\documentclass[showkeys,superscriptaddress,floatfix,prd,10pt,aps]{revtex4-2}
\usepackage{graphicx,epstopdf}
\usepackage[caption=false]{subfig}
\usepackage{appendix}
\usepackage[T1]{fontenc}
\usepackage{lmodern}
\usepackage[dvipsnames,x11names]{xcolor}
\usepackage[colorlinks=true,linkcolor=NavyBlue,citecolor=ForestGreen,urlcolor=NavyBlue]{hyperref}
\usepackage[sort&compress]{natbib}
\usepackage{lipsum}
\usepackage{morefloats}
\usepackage[pdf]{pstricks}
\usepackage{amsmath}
\usepackage{amssymb}
\usepackage{amsfonts}
\usepackage{rotating}
\usepackage{cancel}
\usepackage{mathtools}
\usepackage{bbm}
\usepackage{dsfont}
\usepackage{bbold}
\usepackage{multirow}
\usepackage{ulem}
\usepackage{physics}
\usepackage{orcidlink}
\usepackage{colortbl}

\def\pslash{p\!\!\!\slash }
\def\p_1slash{p1\!\!\!\slash }
\def\p_2slash{p2\!\!\!\slash }
\def\qslash{q\!\!\!\slash }
\def\xslash{x\!\!\!\slash }
\def\eslash{\varepsilon\!\!\!\slash }

\def\vel{\left|}
\def\ver{\right|}

\begin{document}

\title{Unveiling the electromagnetic structure and intrinsic dynamics of spin-$\frac{3}{2}$ hidden-charm pentaquarks: A comprehensive QCD analysis}

\author{Ula\c{s}~\"{O}zdem\orcidlink{0000-0002-1907-2894}}%
\email[]{ulasozdem@aydin.edu.tr }
\affiliation{ Health Services Vocational School of Higher Education, Istanbul Aydin University, Sefakoy-Kucukcekmece, 34295 Istanbul, T\"{u}rkiye}

\begin{abstract}

In this study, we investigate the electromagnetic properties$-$ specifically, the magnetic dipole, electric quadrupole, and magnetic octupole moments$-$ of six hidden-charm pentaquark states: $[u u][d c] \bar c$, $[dd][u c] \bar c$, $[u u][s c] \bar c$, $[dd] [s c] \bar c$, $[s s][u c] \bar c$, and $[s s][d c] \bar c$. Employing the framework of QCD light-cone sum rules and utilizing two distinct diquark-diquark-antiquark interpolating currents, we focus on pentaquark configurations with spin-parity quantum numbers $\mathrm{J^P =\frac{3}{2}^-}$.  From the numerical results, we observe significant deviations between the magnetic dipole moment predictions obtained using different diquark-diquark-antidiquark structures. These results suggest that multiple pentaquark states with identical quantum numbers and quark constituents may exhibit distinct magnetic dipole moments, depending on their internal quark configurations. The obtained electromagnetic moments, particularly the variations in magnetic dipole moments, may provide insights into the internal structure of hidden-charm pentaquark states. 
\end{abstract}
%\keywords{Electromagnetic properties, diquark-diquark-antiquark configuration, QCD light-cone sum rules}

\maketitle

\section{Introduction}\label{motivation}

The study of multiquark states, including tetraquarks, hybrids, glueballs, and pentaquarks, along with the conventional meson-baryon states, has emerged as a pivotal subject in the field of hadron physics after their introduction within the framework of the quark model. Given that Quantum Chromodynamics (QCD) does not inherently preclude their existence, these multiquark states have garnered the interest of the hadron physics community from its inception and have been the subject of comprehensive investigation over an extended period. Following an extended period of anticipation, the much-speculated confirmation of the existence of these states was finally established in 2003 with the announcement by the Belle Collaboration of the discovery of the X(3872) particle \cite{Belle:2003nnu}. This discovery was identified as a candidate tetraquark state, signifying a substantial advancement in the domain of particle physics. Subsequently, the observation of a large number of promising candidates for multi-quark states by different experimental collaborations contributed to the advancement of our expectations and understanding in the field of particle physics. %Subsequent theoretical investigations have sought to elucidate the internal structure and quantum numbers of these particles, which remain ambiguous and require further elucidation through additional research.  The study of multiquark states in particle physics remains a dynamic and captivating area of research, with ongoing efforts to comprehend their fundamental nature~\cite{Esposito:2014rxa, Esposito:2016noz, Olsen:2017bmm, Lebed:2016hpi, Nielsen:2009uh, Brambilla:2019esw, Agaev:2020zad, Chen:2016qju, Ali:2017jda, Guo:2017jvc, Liu:2019zoy, Yang:2020atz, Dong:2021juy, Dong:2021bvy, Chen:2022asf, Meng:2022ozq, Liu:2024uxn, Wang:2025sic}.  However, the fundamental characteristics of these systems are yet to be fully elucidated, necessitating further exploration to attain a comprehensive understanding.

Following the 2003 discovery, another breakthrough occurred in 2015 with the observation by the LHCb Collaboration of a new member of the exotic states, the pentaquark state, which consists of five valence quarks. A detailed analysis of the $J /\psi + p$ decay channel revealed two pentaquark candidates, designated as $P_c(4380)$ and $P_c(4450)$ \cite{LHCb:2015yax}.  In 2019, a more extensive analysis was conducted using a larger data sample, providing further information. It was reported that the previously identified $P_c(4450)$ pentaquark candidate splits into $P_c(4440)$ and $P_c(4457)$ states, revealing an additional peak: the $P_c(4312)$ state \cite{LHCb:2019kea}. It is noteworthy that the final status of the $P_c(4380)$ pentaquark, as reported in the 2015 analysis, remains unresolved, neither confirmed nor ruled out in the updated analysis of 2019. 
In 2020, the LHCb Collaboration unveiled an additional pentaquark state, designated as $P_{cs} (4459)$, within the invariant mass spectrum of $J /\psi\Lambda$ \cite{LHCb:2020jpq}. In 2022, the LHCb collaboration made a notable observation: a novel structure, $P_{cs} (4338)$, was identified within the $J /\psi\Lambda$ mass distribution in the $B^- \to J /\psi\Lambda^- p$ decays \cite{LHCb:2022ogu}. 
In 2024, the Belle Collaboration used $\Upsilon(1S,2S)$ events to search for the pentaquark state in the $pJ/\psi$ final state. No significant $P_{c} (4312)$, $P_{c} (4440)$, or $P_{c} (4457)$ signal was found in the $pJ/\psi$ final state in $\Upsilon(1S,2S)$ inclusive decays \cite{Belle:2024mcb}.
Very recently, Belle Collaboration found evidence of the $P_{cs}(4459)$ state with a significance of 3.3 standard deviations,    including statistical and systematic uncertainties \cite{Belle:2025pey}.  
A compendium of data about the masses, widths, minimal valence quark contents, and observed channels for these states is provided in Table \ref{pentaquarks}.  Along with the aforementioned hidden-charm pentaquark states, searches for doubly and triply strange hidden-charm pentaquarks are currently underway, with the CMS Collaboration recently observing the decay $\Lambda_b^0\to J/\psi\Xi^-K^+$~\cite{CMS:2024vnm}. Nevertheless, insufficient yield and inadequate resolution prevented the observation of a clear spectrum in the $J/\psi\Xi^-$ invariant mass.  These findings are relevant for illuminating the strong interaction processes that underlie the hadronic decays of beauty baryons and the underlying potential mechanism for the production of multiquark states.
\begin{widetext}

\begin{table}[htp]
\caption{Reported hidden-charm pentaquark states by the LHCb \cite{LHCb:2015yax, LHCb:2019kea, LHCb:2020jpq, LHCb:2022ogu} and Belle \cite{Belle:2025pey} collaborations.}\label{pentaquarks}
\begin{tabular}{l|c|c|c|c}
\toprule %[0.8pt] 
%\noalign{\smallskip}
State  & Mass (MeV) & Width (MeV) & Quark content &  Observed decays \\
\toprule%[0.8pt] 
%\noalign{\smallskip}
%$P_c(4380)^+$ \cite{LHCb:2015yax}  &            $4380\pm8\pm29$          &       $215\pm18\pm86$           & $uudc\bar{c}$ & $\Lambda_b^0 \to J/\psi pK^-$\\
%\noalign{\smallskip}
$P_c(4312)^+$ \cite{LHCb:2019kea}       & ~~$4311.9\pm0.7^{~+6.8}_{~-0.6}$~~  &  $9.8\pm2.7^{~+3.7}_{~-4.5}$    & $uudc\bar{c}$ & $\Lambda_b^0 \to J/\psi pK^-$ \\
%\noalign{\smallskip}
$P_c(4440)^+$ \cite{LHCb:2019kea}      &    $4440.3\pm1.3^{~+4.1}_{~-4.7}$   &  $20.6\pm4.9^{~+8.7}_{~-10.1}$  & $uudc\bar{c}$& $\Lambda_b^0 \to J/\psi pK^-$ \\
%\noalign{\smallskip}
$P_c(4457)^+$  \cite{LHCb:2019kea}    &    $4457.3\pm0.6^{~+4.1}_{~-1.7}$   &  $6.4\pm2.0^{~+5.7}_{~-1.9}$    & $uudc\bar{c}$ & $\Lambda_b^0 \to J/\psi pK^-$ \\
%\noalign{\smallskip}
$P_{cs}(4459)^0$ \cite{LHCb:2020jpq}   &    $4458.8\pm2.9^{~+4.7}_{~-1.1}$   &  $17.3\pm6.5^{~+8.0}_{~-5.7}$   & $udsc\bar{c}$ &~$\Xi_b^- \to J/\psi \Lambda K^-$ \\
%\noalign{\smallskip}
$P_{cs}(4338)^0$ \cite{LHCb:2022ogu}  &    $4338.2 \pm 0.7 \pm 0.4$   &  $7.0 \pm 1.2 \pm 1.3$    & $udsc\bar{c}$ & $B^-\to J/\psi \Lambda \bar{p}$\\
%\noalign{\smallskip}
$P_{cs}(4459)^0$ \cite{Belle:2025pey}   &    $4471.7 \pm 4.8 \pm 0.6$   &  $21.9 \pm 13.1 \pm 2.7$   & $udsc\bar{c}$ &~$\Upsilon (1S, 2S)\to J/\psi \Lambda $ \\
\toprule%[0.8pt]
\end{tabular}
\end{table}

\end{widetext}

Several interpretations have been posited for these pentaquark states, including the hypothesis of tightly-bound pentaquark states, loosely-bound meson-baryon molecular states, and the product of re-scattering effects, among others. For a more detailed discussion, see Refs.~\cite{Esposito:2014rxa, Esposito:2016noz, Olsen:2017bmm, Lebed:2016hpi, Nielsen:2009uh, Brambilla:2019esw, Agaev:2020zad, Chen:2016qju, Ali:2017jda, Guo:2017jvc, Liu:2019zoy, Yang:2020atz, Dong:2021juy, Dong:2021bvy, Chen:2022asf, Meng:2022ozq, Liu:2024uxn, Wang:2025sic}. 
Despite the significant body of research conducted since the initial observation of these states, our understanding of their precise nature, internal structure, and crucial parameters such as spin-parity quantum numbers remains incomplete and requires further elucidation. A considerable amount of effort is still required to address the aforementioned inquiries. 
In addition to the discovered pentaquark states $P_c(4312)$, $P_c(4440)$, $P_c(4457)$, $P_{cs}(4338)$, and $P_{cs}(4459)$, searching for other possible pentaquarks containing one or two strange quarks, or without strangeness, and determining their properties are also important and interesting topics of study.
One of the key tools for probing the internal structure of hadrons is the study of their electromagnetic properties, such as magnetic dipole moments, electric quadrupole moments, and higher-order multipole moments. These properties are sensitive to the spatial distribution of quarks and their spin orientations within the hadron. For hidden-charm pentaquarks, the electromagnetic multipole moments can provide critical insights into their quark-gluon composition, spin-parity assignments, and geometric shape. To this end, the present study investigates the magnetic dipole, electric quadrupole, and magnetic octupole moments of the $[u u][d c] \bar c$, $[dd][u c] \bar c$, $[u u][s c] \bar c$, $[dd] [s c] \bar c$, $[s s][u c] \bar c$, and $[s s][d c] \bar c$ states in the context of the QCD light-cone sum rules~\cite{Chernyak:1990ag, Braun:1988qv, Balitsky:1989ry}. In the course of examining these properties, two distinct diquark-diquark-antiquark forms of the interpolating currents are employed, given that these pentaquark candidates possess quantum numbers $\mathrm{J^P =\frac{3}{2}^-}$. By calculating the magnetic dipole, electric quadrupole, and magnetic octupole moments, we aim to shed light on the internal structure of these exotic states and provide predictions that can be tested in future experiments. Our results not only contribute to the ongoing effort to understand the nature of pentaquarks but also offer valuable insights into the dynamics of multiquark systems in QCD.  Despite their importance, the electromagnetic properties of hidden-charm pentaquarks have not been extensively studied, either theoretically or experimentally. There is a paucity of research in the extant literature on the electromagnetic multipole moments of hidden-charm/bottom pentaquarks~\cite{Wang:2016dzu, Ozdem:2018qeh, Ortiz-Pacheco:2018ccl, Xu:2020flp, Ozdem:2021btf, Ozdem:2021ugy, Li:2021ryu, Ozdem:2023htj, Wang:2023iox, Ozdem:2022kei, Gao:2021hmv, Guo:2023fih, Ozdem:2022iqk, Wang:2022nqs, Wang:2022tib, Ozdem:2024jty, Li:2024wxr, Li:2024jlq,  Ozdem:2024rqx, OZDEM:2024jlw, Mutuk:2024ltc, Mutuk:2024jxf, Mutuk:2024ach}.

The paper is organized as follows: Sec.~\ref{formalism} outlines the QCD light-cone sum rules utilized to calculate the magnetic dipole, electric quadrupole, and magnetic octupole moments of the states being studied.
In Sec.~\ref{numerical}, we present the numerical analysis of the results derived from the QCD light-cone sum rules. The final section offers a summary and conclusion of the findings.

\begin{widetext}

\section{ Formalism}\label{formalism}

To facilitate the analysis of the electromagnetic multipole moments of the $P_{c(s)}$ states, it is necessary to determine the correlation function that will be employed in this analysis. The following formula has been determined to be the correlation function that should be used: 
\begin{eqnarray} \label{edmn01}
\Pi_{\mu \nu}(p,q)&=&i\int d^4x e^{ip \cdot x} \langle0|T\left\{J_{\mu}^i(x)\bar{J}_{\nu}^i(0)\right\}|0\rangle _\gamma \, ,
\end{eqnarray}
where  the $J_{\mu}^i(x)$ stand for interpolating current of the $P_{c(s)}$ states,  $\gamma$ is the external electromagnetic field, and $q$ is the momentum of the photon.   To study the electromagnetic multipole moments of $P_{c(s)}$ pentaquarks, the $J_{\mu}^i(x)$ interpolating currents prove to be significant and can be considered pivotal parameters. In consideration of the quark contents and spin-parity quantum numbers of the pentaquarks in question, the interpolating currents that are likely to couple to these pentaquarks with spin-parity quantum numbers $\mathrm{J^P =\frac{3}{2}^-}$ are expressed as follows \cite{Wang:2015ixb}:

\begin{eqnarray}
J_{\mu}^1(x)&=&\frac{\mathcal A }{\sqrt{3}} \bigg\{ \big[ {q_1}^T_d(x) C \gamma_\mu {q_1}_e(x) \big] \big[ {q_2}^T_f(x) C \gamma_\alpha c_g(x)\big] + 2
\big[ {q_1}^T_d(x) C \gamma_\mu {q_2}_e(x) \big] \big[ {q_1}^T_f(x) C \gamma_\alpha c_g(x)\big]  \bigg\}  \gamma_5 \gamma^\alpha C  \bar{c}^{T}_{c}(x)\, , %\\
%%%%%%%%%%%%%%%%%%%%%%%%%%%%%%%%%%%%%%%%%%%%%
\end{eqnarray}
\begin{eqnarray}
%%%%%%%%%%%%
J_{\mu}^2(x)&=&\frac{\mathcal A }{\sqrt{3}} \bigg\{ \big[ {q_1}^T_d(x) C \gamma_\alpha {q_1}_e(x) \big] \big[ {q_2}^T_f(x) C \gamma_\mu c_g(x)\big] + 2
\big[ {q_1}^T_d(x) C \gamma_\alpha {q_2}_e(x) \big] \big[ {q_1}^T_f(x) C \gamma_\mu c_g(x)\big]  \bigg\}  \gamma_5 \gamma^\alpha C  \bar{c}^{T}_{c}(x)\, ,
\end{eqnarray}
where $ \mathcal A = \varepsilon_{abc}\varepsilon_{ade} \varepsilon_{bfg}$ with   
$a$, $b$, $c$, $d$, $e$, $f$ and $g$ being color indices; and the $C$ is the charge conjugation operator. The quark content of the $P_{c(s)}$ states is listed in Table \ref{quarkcon}. 
We would like to note that the interpolating currents considered here may also couple to positive-parity pentaquark states. According to Refs. \cite{Wang:2015ixb, Duan:2024uuf, Wang:2015wsa, Wang:2015ava}, pentaquarks with positive parity have higher masses than those with negative parity states. This mass difference is particularly relevant in magnetic moment calculations, where contributions from higher-mass states are typically neglected due to their suppressed effects. 
Since the pentaquarks studied in this work are low-mass states, it is reasonable to disregard the already small contribution of the spin-$\frac{3}{2}^+$ pentaquark states in the magnetic moment analysis. Therefore, the interpolating current used predominantly couples to the spin-$\frac{3}{2}^-$ pentaquarks, and the resulting magnetic moment values correspond primarily to these states.
\begin{table}[htp]
	\addtolength{\tabcolsep}{10pt}
		\begin{center}
		\caption{The quark content of the $P_{c(s)}$ states.}
	\label{quarkcon}
\begin{tabular}{lccccccccc}
	   \hline\hline
	 %  \\
  States& $[u u][d c] \bar c$&$[dd][u c] \bar c$&$[u u][s c] \bar c$&$[dd ][s c] \bar c$&$[s s][u c] \bar c$&$[s s][d c] \bar c$\\
  %\\
\hline\hline
%\\
$q_1$&u&d& u&d&s&s\\
%\\
$q_2$&d&u& s&s&u&d\\
%\\
	   \hline\hline
\end{tabular}
\end{center}
\end{table}

The essential parameters necessary to initiate the analysis are derived within the framework of the pertinent methodology. As a preliminary step, we will derive some of the analytical expressions for the electromagnetic multipole moments using the hadronic description of the correlation function.

\subsection{Hadronic description}

In the context of the hadron description, a complete set of intermediate hadronic states is inserted into the correlation function, each state being assigned the same quantum numbers as the interpolating currents. The ground-state contributions are then isolated. Consequently, we get 

\begin{eqnarray}\label{edmn02}
\Pi^{Had}_{\mu\nu}(p,q)&=&\frac{\langle0\mid J_{\mu}(x)\mid
P_{c(s)}(p_2,s)\rangle}{[p_2^{2}-m_{P_{c(s)}}^{2}]}\langle P_{c(s)}(p_2,s)\mid
P_{c(s)}(p_1,s)\rangle_\gamma\frac{\langle P_{c(s)}(p_1,s)\mid
\bar{J}_{\nu}(0)\mid 0\rangle}{[p_1^{2}-m_{P_{c(s)}}^{2}]},
\end{eqnarray}
where $p_1 = p+q$, $p_2=p$. Upon examination of the aforementioned equation, it becomes evident that a variety of matrix elements of differing structures are necessary for the subsequent analysis. These matrix elements are outlined below~\cite{Weber:1978dh,Nozawa:1990gt,Pascalutsa:2006up,Ramalho:2009vc}:
\begin{eqnarray}
\label{lambdabey}
\langle0\mid J_{\mu}(x)\mid P_{c(s)}(p_2,s)\rangle &=&\lambda_{P_{c(s)}}u_{\mu}(p_2,s),\\
%\nonumber\\
%%%%%%%%%%%%%%%%%%%%%%%%%%%%%%%%%%%%%%%%%%%%%%%%%%%%%%%%%%%%%%%%%%%%
\langle {P_{c(s)}}(p_1,s)\mid
\bar{J}_{\nu}(0)\mid 0\rangle &=& \lambda_{{P_{c(s)}}}\bar u_{\nu}(p_1,s),\\
%\nonumber\\
%%%%%%%%%%%%%%%%%%%%%%%%%%%%%%%%%%%%%%%%
\langle P_{c(s)}(p_2,s)\mid P_{c(s)}(p_1,s)\rangle_\gamma &=&-e \,\bar
u_{\mu}(p_2,s)\left\{F_{1}(q^2)g_{\mu\nu}\eslash-
\frac{1}{2m_{P_{c(s)}}}\left
[F_{2}(q^2)g_{\mu\nu} \eslash\qslash+F_{4}(q^2)\frac{q_{\mu}q_{\nu} \eslash\qslash}{(2m_{P_{c(s)}})^2}\right]
\right.\nonumber\\&+&\left.
\frac{F_{3}(q^2)}{(2m_{P_{c(s)}})^2}q_{\mu}q_{\nu}\eslash\right\} u_{\nu}(p_1,s),\label{matelpar}
\end{eqnarray}
where $\lambda_{P_{c(s)}}$  is pole residue (or coupling) of the $P_{c(s)}$ states; $u_{\mu}(p_2,s)$ and $ \bar u_{\nu}(p_1,s)$ are the spinors of the $P_{c(s)}$ states; $\varepsilon$ is the photon's polarization vector, and $F_i (q^2)$ are transition form factors.  By employing the Eqs.~(\ref{edmn02}) to (\ref{matelpar}) and making the requisite simplifications, the expressions for the magnetic moment of the $P_{c(s)}$ states in conjunction with hadronic parameters are as follows:
\begin{eqnarray}
\label{final phenpart}
\Pi^{Had}_{\mu\nu}(p,q)&=&\frac{\lambda_{_{P_{c(s)}}}^{2}}{[(p+q)^{2}-m_{_{P_{c(s)}}}^{2}][p^{2}-m_{_{P_{c(s)}}}^{2}]}
\bigg[  g_{\mu\nu}\pslash\eslash\qslash \,F_{1}(q^2) 
-m_{P_{c(s)}}g_{\mu\nu}\eslash\qslash\,F_{2}(q^2)+
\frac{F_{3}(q^2)}{2m_{P_{c(s)}}}q_{\mu}q_{\nu}\eslash\qslash\, \nonumber\\&+&
\frac{F_{4}(q^2)}{4m_{P_{c(s)}}^3}(\varepsilon.p)q_{\mu}q_{\nu}\pslash\qslash \,+
\mathrm{other~independent~structures} \bigg].
\end{eqnarray}
To derive the aforementioned equation, it is necessary to perform a summation over the spins of the $P_c(s)$ state in the following manner: 
\begin{align}\label{raritabela}
\sum_{s}u_{\mu}(p,s)\bar u_{\nu}(p,s)=-\big(\pslash+m_{P_{c(s)}}\big)\Big[g_{\mu\nu}
-\frac{1}{3}\gamma_{\mu}\gamma_{\nu}-\frac{2\,p_{\mu}p_{\nu}}
{3\,m^{2}_{P_{c(s)}}}+\frac{p_{\mu}\gamma_{\nu}-p_{\nu}\gamma_{\mu}}{3\,m_{P_{c(s)}}}\Big].
\end{align}

The Lorentz invariant form factors $F_{1}(q^2)$, $F_{2}(q^2)$, $F_{3}(q^2)$ and $F_{4}(q^2)$ are related to the magnetic dipole, $G_{M}(q^2)$, electric quadrupole, $G_{Q}(q^2)$, and magnetic octupole, $G_{O}(q^2)$, form factors as~\cite{Weber:1978dh,Nozawa:1990gt,Pascalutsa:2006up,Ramalho:2009vc}:
\begin{eqnarray}
G_{M}(q^2) &=& \left[ F_1(q^2) + F_2(q^2)\right] ( 1+ \frac{4}{5}
\tau ) -\frac{2}{5} \left[ F_3(q^2)  +
F_4(q^2)\right] \tau \left( 1 + \tau \right), \nonumber\\
G_{Q}(q^2) &=& \left[ F_1(q^2) -\tau F_2(q^2) \right]  -
\frac{1}{2}\left[ F_3(q^2) -\tau F_4(q^2)
\right] \left( 1+ \tau \right),  \nonumber \\
 G_{O}(q^2) &=&
\left[ F_1(q^2) + F_2(q^2)\right] -\frac{1}{2} \left[ F_3(q^2)  +
F_4(q^2)\right] \left( 1 + \tau \right),\end{eqnarray}
  where $\tau
= -\frac{q^2}{4m^2_{P_{c(s)}}}$ is a kinematic factor. These expressions enable the extraction of the electromagnetic form factors for these pentaquark states. 
At zero momentum transfer, since we deal with the real photon ($q^2 =0$), these form factors
are proportional to the usual static quantities of the magnetic dipole ($\mu_{P_{c(s)}}$), the electric quadrupole
($Q_{P_{c(s)}}$)  and the magnetic octupole ($O_{P_{c(s)}}$) moments of the $P_{c(s)}$ states,
 \begin{eqnarray}\label{mqo2}
\mu_{P_{c(s)}}&=&\frac{e}{2m_{P_{c(s)}}}G_{M}(0),~~~~~~\nonumber\\
Q_{P_{c(s)}}&=&\frac{e}{m_{P_{c(s)}}^2}G_{Q}(0), ~~~~~~ \nonumber\\
O_{P_{c(s)}}&=&\frac{e}{2m_{P_{c(s)}}^3}G_{O}(0),
\end{eqnarray}
 where
 \begin{eqnarray}\label{mqo1}
G_{M}(0)&=&F_{1}(0)+F_{2}(0),\nonumber\\
G_{Q}(0)&=&F_{1}(0)-\frac{1}{2}F_{3}(0),\nonumber\\
G_{O}(0)&=&F_{1}(0)+F_{2}(0)-\frac{1}{2}[F_{3}(0)+F_{4}(0)].
\end{eqnarray}

Along with the derivation of the above expressions, the hadronic description of the correlation function corresponding to the magnetic dipole, electric quadrupole, and magnetic octupole moments of the pentaquark states has been obtained.

 \subsection{QCD description}
In the following, we will provide a concise overview of the operator product expansion for the correlation function in perturbative QCD. To that end, the quark fields in the correlation function are contracted through the use of the Wick theorem.
%%%%%%%%%%%%%
Using the $J_\mu^1$ current as an example, this would result in the following.
\begin{align}
\label{QCD1}
\Pi^{QCD-J_\mu^1}_{\mu\nu}(p,q)&= \frac{i}{3}\,\varepsilon^{abc}\varepsilon^{a^{\prime}b^{\prime}c^{\prime}}\varepsilon^{ade} \varepsilon^{a^{\prime}d^{\prime}e^{\prime}}\varepsilon^{bfg} \varepsilon^{b^{\prime}f^{\prime}g^{\prime}} 
%\mathcal {A} \mathcal{A^\prime}
\int d^4x e^{ip\cdot x}\langle 0|
%%%%%%%%%%%%%%%%%%%%%%%%%%%%%%%%%%%%%%%%%%%%%%%%%%%%%%%%%%%%%%%%%%%%%%%%%%%%%%%%%
\Big\{
 \nonumber\\
&
- \mbox{Tr}\Big[  \gamma_\mu S_{q_1}^{ee^\prime}(x) \gamma_\nu C   S_{q_1}^{dd^\prime \mathrm{T}}(x) C\Big]
 \mbox{Tr}\Big[ \gamma_\alpha S_c^{gg^\prime}(x) \gamma_\beta C  S_{q_2}^{ff^\prime \mathrm{T}}(x)C \Big] 
 %%%%%%%%%%%%%%%%%%%%%%%%%%%%%%%%%%%%%%%%%%%%%%%%%%%%%%%%%%%%%%%%%%%%%%%%%%%%%%%%%
 \nonumber\\
&
+ \mbox{Tr}\Big[  \gamma_\mu S_{q_1}^{ed^\prime}(x) \gamma_\nu C  S_{q_1}^{de^\prime \mathrm{T}}(x) C\Big]
 \mbox{Tr}\Big[ \gamma_\alpha S_c^{gg^\prime}(x) \gamma_\beta C  S_{q_2}^{ff^\prime \mathrm{T}}(x)C  \Big] 
%%%%%%%%%%%%%%%%%%%%%%%%%%%%%%%%%%%%%%%%%%%%%%%%%%%%%%%%%%%%%%%%%%%%%%%%%%%%%%%
\nonumber\\
&
-4 \mbox{Tr}\Big[  \gamma_\mu S_{q_2}^{ee^\prime}(x) \gamma_\nu   C S_{q_1}^{dd^\prime \mathrm{T}}(x) C\Big]
 \mbox{Tr}\Big[ \gamma_\alpha S_c^{gg^\prime}(x) \gamma_\beta C  S_{q_1}^{ff^\prime \mathrm{T}}(x)C \Big] 
 %%%%%%%%%%%%%%%%%%%%%%%%%%%%%%%%%%%%%%%%%%%%%%%%%%%%%%%%%%%%%%%%%%%%%%%%%%%%%%%
 \nonumber\\
 &
+4 \mbox{Tr}\Big[  \gamma_\mu S_{q_2}^{ee^\prime}(x) \gamma_\nu C  S_{q_1}^{fd^\prime \mathrm{T}}(x) C\Big]
 \mbox{Tr}\Big[ \gamma_\alpha S_c^{gg^\prime}(x) \gamma_\beta C  S_{q_1}^{df^\prime \mathrm{T}}(x)C  \Big] 
 %%%%%%%%%%%%%%%%%%%%%%%%%%%%%%%%%%%%%%%%%%%%%%%%%%%%%%%%%%%%%%%%
 \nonumber\\
&
 +2 \mbox{Tr} \Big[ \gamma_\alpha S_c^{gg^\prime}(x) 
\gamma_\beta C S_{q_1}^{ef^\prime \mathrm{T}}(x)  C \gamma_\mu S_{q_1}^{dd^\prime}(x) \gamma_\nu C  S_{q_2}^{fe^\prime \mathrm{T}}(x) C\Big]
%%%%%%%%%%%%%%%%%%%%%%%%%%%%%%%%%%%%%%%%%%%%%%%%%%%%%%%%%%%%%%%%%%%%%%%%%%%%%%%
 %%%%%%%%%%%%%%%%%
 \nonumber\\
&
 -2 \mbox{Tr} \Big[ \gamma_\alpha S_c^{gg^\prime}(x) 
\gamma_\beta C S_{q_1}^{df^\prime \mathrm{T}}(x)  C \gamma_\mu S_{q_1}^{ed^\prime}(x) \gamma_\nu C  S_{q_2}^{fe^\prime \mathrm{T}}(x) C\Big]
\nonumber\\
%%%%%%%%%%%%%%%%%%%%%%%%%%%%%%%%%%%%%%%%%%%%%%%%%%%%%%%%%%%%%%%%%%%%%%%%%%%%%%%
&
+2 \mbox{Tr} \Big[ \gamma_\alpha S_c^{gg^\prime}(x) 
\gamma_\beta C S_{q_2}^{ef^\prime \mathrm{T}}(x)  C \gamma_\mu S_{q_1}^{dd^\prime}(x) \gamma_\nu C  S_{q_1}^{fe^\prime \mathrm{T}}(x) C\Big]
%%%%%%%%%%%%%%%%%%%%%%%%%%%%%%%%%%%%%%%%%%%%%%%%%%%%%%%%%%%%%%%%%%%%%%%%%%%%%%%
\nonumber\\
&
 -2 \mbox{Tr} \Big[ \gamma_\alpha S_c^{gg^\prime}(x) 
\gamma_\beta C S_{q_2}^{ef^\prime \mathrm{T}}(x)  C \gamma_\mu S_{q_1}^{de^\prime}(x) \gamma_\nu C  S_{q_1}^{fd^\prime \mathrm{T}}(x) C\Big]
\Big \} 
\Big(\gamma_5 \gamma_\alpha C S_c^{c^{\prime}c \mathrm{T}} (-x) C \gamma_\beta \gamma_5\Big)
|0 \rangle_\gamma ,
\end{align}
where  the light and charm quark propagators, denoted by $S_q(x)$ and $S_c(x)$, respectively, can be expressed as follows:~\cite{Balitsky:1987bk, Belyaev:1985wza}
\begin{align}
\label{edmn13}
S_{q}(x)&= S_q^{free}(x) - \frac{\langle \bar qq \rangle }{12} \Big(1-i\frac{m_{q} \xslash}{4}   \Big)- \frac{ \langle \bar qq \rangle }{192}
m_0^2 x^2  \Big(1-i\frac{m_{q} \xslash}{6}   \Big)
-\frac {i g_s }{16 \pi^2 x^2} \int_0^1 du \, G^{\mu \nu} (ux)
\bigg[\bar u \rlap/{x} 
\sigma_{\mu \nu} + u \sigma_{\mu \nu} \rlap/{x}
 \bigg],\\
%\nonumber\\
%\end{align}%
%and
%
%\begin{align}
S_{Q}(x)&=S_Q^{free}(x)
%\nonumber\\
%&
-i\frac{m_{Q}\,g_{s} }{16\pi ^{2}}  \int_0^1 dv \,G^{\mu \nu}(vx)\bigg[ (\sigma _{\mu \nu }{\xslash}
 % \nonumber\\
%&
+{\xslash}\sigma _{\mu \nu }) 
    \frac{K_{1}\big( m_{Q}\sqrt{-x^{2}}\big) }{\sqrt{-x^{2}}}
   %\nonumber\\
  %&
 +2\sigma_{\mu \nu }K_{0}\big( m_{Q}\sqrt{-x^{2}}\big)\bigg],
 \label{edmn14}
\end{align}%
with  
\begin{align}
 S_q^{free}(x)&=\frac{1}{2 \pi x^2}\Big(i \frac{\xslash}{x^2}- \frac{m_q}{2}\Big),\\
 %\end{align}
 %\begin{align}
 S_c^{free}(x)&=\frac{m_{c}^{2}}{4 \pi^{2}} \bigg[ \frac{K_{1}\big(m_{c}\sqrt{-x^{2}}\big) }{\sqrt{-x^{2}}}
+i\frac{{\xslash}~K_{2}\big( m_{c}\sqrt{-x^{2}}\big)}
{(\sqrt{-x^{2}})^{2}}\bigg],
\end{align}
where $G^{\mu\nu}$ is the gluon field strength tensor and 
$K_n$ are modified Bessel functions of the second kind. %In this work, we employ the integral representation of the modified Bessel function of the second kind as follows:     
%\begin{equation}\label{b2}
%K_n(m_Q\sqrt{-x^2})=\frac{\Gamma(n+ 1/2)~2^n}{m_Q^n \,\sqrt{\pi}}\int_0^\infty dt~\cos(m_Qt)\frac{(\sqrt{-x^2})^n}{(t^2-x^2)^{n+1/2}}.
%\end{equation}

In the QCD description, there are two classes of contributions according to the way the photon interacts with the quark lines. First, the photon interacts perturbatively with the quark line through the standard QED interaction.  In this set of interactions, it is possible to replace one of the free quark propagators in Eq. (\ref{QCD1}) with a propagator that incorporates an electromagnetic interaction. This replacement can be achieved by employing the following method: 
\begin{align}
\label{free}
S^{free}(x) \rightarrow \int d^4z\, S^{free} (x-z)\,\rlap/{\!A}(z)\, S^{free} (z)\,.
\end{align}
%where the four surviving propagators are regarded as full quark propagators. 

The second category of interactions encompasses the non-perturbative interaction of photons with quarks, as depicted by the photon light cone distribution amplitude. One of the five propagators in Eq. (\ref{QCD1}) is replaced by
\begin{align}
\label{edmn15}
S_{\alpha\beta}^{ab}(x) \rightarrow -\frac{1}{4} \big[\bar{q}^a(x) \Gamma_i q^b(0)\big]\big(\Gamma_i\big)_{\alpha\beta},
\end{align}
where   $\Gamma_i = \mathrm{1}, \gamma_5, \gamma_\mu, i\gamma_5 \gamma_\mu, \sigma_{\mu\nu}/2$. Upon substituting the replacement stipulated in Eq.(\ref{edmn15}), expressions such as $\langle \gamma(q)\vel \bar{q}(x) \Gamma_i G_{\mu\nu}q(0) \ver 0\rangle$  and $\langle \gamma(q)\vel \bar{q}(x) \Gamma_i q(0) \ver 0\rangle$ are derived. These expressions are obtained in the form of photon dispersion amplitudes within the framework of the requisite analysis (for details see Ref.~\cite{Ball:2002ps}).    
As these subjects are thoroughly delineated and have been standardized, we have elected to omit further elaboration in this section. Individuals seeking a more comprehensive understanding of these subjects are encouraged to refer to the Refs.~\cite{Ozdem:2022vip, Ozdem:2022eds}.  Eqs.~(\ref{free}) and ~(\ref{edmn15}) have been utilized to encompass both perturbative and non-perturbative contributions within the conducted analysis, under the prescribed scheme.

\subsection{QCD light-cone sum rules for the electromagnetic multipole moments}

The findings from evaluations on both descriptions of the correlation function are compared by using dispersion relations that take into account the coefficients of the same Lorentz structures. In the final step, double Borel transformations are applied to the variables $-p^2$ and $-(p+q)^2$. This process eliminates contributions from the continuum and higher states, thereby enhancing the contributions from ground states. %To illustrate, consider the magnetic dipole moment, for which the results are as follows via double dispersion relations \cite{Aliev:2023pwd}: 
To illustrate this, consider the case of the magnetic dipole moment. The results obtained using the double dispersion relations are as follows \cite{Ioffe:1983ju, Aliev:2023pwd}:
\begin{equation}
    \begin{aligned}
                \mu_{P_{c(s)}} \lambda^2_{P_{c(s)}} e^{-\frac{m_1^2}{M_1^2}}e^{-\frac{m_2^2}{M_2^2}} + \cdots  &= \int_0^\infty ds_1 \int_0^\infty ds_2 \, e^{-\frac{s_1}{M_1^2}-\frac{s_2}{M_2^2}}\rho(s_1,s_2),\\
    \end{aligned}
    \label{correlated}
\end{equation}
where $m_1(m_2)$, $s_1(s_2)$ and $M_1^2(M_2^2)$ are the mass, continuum threshold, and Borel parameter for the initial (final) $P_{c(s)}$ tetraquarks, respectively,   and $\cdots$ denote
the contribution from higher states and the continuum. The $\rho(s_1, s_2)$ denotes the hadronic spectral density, which is obtained through double Borel transformations applied to the correlation function.

 To calculate the magnetic dipole moment within the QCD light-cone sum rules, it is necessary to extract the contributions from higher states and the continuum. This is achieved by utilizing the quark-hadron duality ansatz as follows:
\begin{equation}
    \rho(s_1,s_2) \simeq \rho^{OPE}(s_1,s_2) ~~\mbox{if}~~ (s_1,s_2) \not\in {\mathbb D},
\end{equation}
where $\mathbb D$ is a domain in the $(s_1,s_2)$ plane. Generally, the domain $\mathbb D$ is a rectangular region defined by $s_1<s_{10}$ and $s_2<s_{20}$ for some constants $s_{10}$ and $s_{20}$, or a triangular region. In this work, for the sake of brevity, the continuum subtraction is carried out by selecting $\mathbb D$ as the region specified as $s \equiv s_1 u_0 + s_2 \bar u_0 < s_0$ where $u_0\equiv \frac{M_2^2}{M_1^2+M_2^2}$ and $\bar u_0 = 1 - u_0$. Defining  a second variable $u=\frac{s_1 u_0}{s}$, the integral in the $(s_1,s_2)$ plane can be defined as:
\begin{equation}
    \int_0^\infty ds_1 \int_0^\infty ds_2 \,e^{-\frac{s_1}{M_1^2}-\frac{s_2}{M_2^2}}\, \rho(s_1,s_2) = \int_0^\infty ds \,\rho(s)\, e^{-\frac{s}{M^2}},
\end{equation}
where   
\begin{equation}
{M^2}= \frac{M_1^2 M_2^2}{M_1^2+M_2^2}, ~\rm{and}  ~
    \rho(s) = \frac{s}{u_0 \bar u_0} \int_0^1 du \,\rho\left( s\frac{u}{u_0}, s \frac{\bar u}{\bar u_0} \right). 
\end{equation}

In the context of the problem being examined, the masses of the initial and final states are the same, so we can set $M_1^2 = M_2^2 = 2M^2$, which gives $u_0=\frac{1}{2}$. Following the conclusion of the aforementioned procedure, the continuum subtraction by setting the upper limit to $s_0$  is equivalent (in the original double spectral density) to subtract everything outside the triangular region $s = s_1 u_0 + s_2 \bar u_0 \equiv s_0$. Following this scheme, the QCD light-cone sum rules for the  $F_1$, $F_2$, $F_3$, and $F_4$ form factors can be determined by equating the coefficients of the $g_{\mu\nu}\pslash\eslash\qslash$, $g_{\mu\nu}\eslash\qslash$, $q_{\mu}q_{\nu}\eslash\qslash$ and  $(\varepsilon.p)q_{\mu}q_{\nu}\pslash\qslash$ structures.   
The obtained sum rules for the electromagnetic multipole moments of $P_{c(s)}$ states are presented as follows,
 \begin{align}
  \label{edmn0015}
&\mu^{J_\mu^1}_{ P_{c(s)}} \,\lambda^2_{ P_{c(s)}}=e^{\frac{m^2_{ P_{c(s)}}}{\rm{M^2}}}\, \rho_1 (\rm{M^2},\rm{s_0}),~~~~~~~~~~~~~~~~
\mu^{J_\mu^2}_{ P_{c(s)}} \,\lambda^2_{ P_{c(s)}}=e^{\frac{m^2_{ P_{c(s)}}}{\rm{M^2}}}\, \rho_4 (\rm{M^2},\rm{s_0}),\\
%%%%%%%%%%%%%%%%%%%%%%%%%%%%%%%%%%%%%%%%%%%%%%%%%%%%%%%%%%%%%%%%%%%%%%%%%%%%%%%%%%%%%%%%%%%%%%%
&\mathcal Q^{J_\mu^1}_{ P_{c(s)}} \,\lambda^2_{ P_{c(s)}}=e^{\frac{m^2_{ P_{c(s)}}}{\rm{M^2}}}\, \rho_2 (\rm{M^2},\rm{s_0}), ~~~~~~~~~~~~~~~~    
\mathcal Q^{J_\mu^2}_{ P_{c(s)}} \,\lambda^2_{ P_{c(s)}}=e^{\frac{m^2_{ P_{c(s)}}}{\rm{M^2}}}\, \rho_5 (\rm{M^2},\rm{s_0}),\\
%%%%%%%%%%%%%%%%%%%%%%%%%%%%%%%%%%%%%%%%%%%%%%%%%%%%%%%%%%%%%%%%%%%%%%%%%%%%%%%%%%%%%%%%%%%%%%%
&\mathcal O^{J_\mu^1}_{ P_{c(s)}} \,\lambda^2_{ P_{c(s)}}=e^{\frac{m^2_{ P_{c(s)}}}{\rm{M^2}}}\, \rho_3 (\rm{M^2},\rm{s_0}), ~~~~~~~~~~~~~~~~
\mathcal O^{J_\mu^2}_{ P_{c(s)}} \,\lambda^2_{ P_{c(s)}}=e^{\frac{m^2_{ P_{c(s)}}}{\rm{M^2}}}\, \rho_6 (\rm{M^2},\rm{s_0}).
 \end{align}
 
 As the forms of the $\rho_i (\rm{M^2},\rm{s_0})$ functions are similar, for illustrative purposes, we only present the explicit form of the $\rho_1 (\rm{M^2},\rm{s_0})$, $\rho_2 (\rm{M^2},\rm{s_0})$ and $\rho_3 (\rm{M^2},\rm{s_0})$ functions in the Appendix.

\end{widetext}

\section{Results and Discussions}\label{numerical}

 The following section is concerned with the results of a numerical analysis of the QCD light-cone sum rules. This analysis is conducted to predict the electromagnetic multipole moments of the $P_{c(s)}$ states.  To perform a numerical analysis of the QCD light-cone sum rule, it is first necessary to ascertain the numerical values of several parameters. The following numerical values are assigned to the relevant parameters: $m_s =93.4^{+8.6}_{-3.4}\,\mbox{MeV}$, $m_c = 1.27 \pm 0.02\,\mbox{GeV}$~\cite{ParticleDataGroup:2022pth}, $m_{[uu][dc]\bar c} =$ $ m_{[dd][uc]\bar c}=$ $  4.39 \pm 0.14$ GeV~\cite{Wang:2015ixb}, $m_{[uu][sc]\bar c} =$ $ m_{[dd][sc]\bar c}=$ $  4.51 \pm 0.12$ GeV~\cite{Wang:2015ixb}, 
$m_{[ss][uc]\bar c} =$ $ m_{[ss][dc]\bar c} =$ $ 4.60 \pm 0.11$ GeV~\cite{Wang:2015ixb},  $\langle \bar uu\rangle = 
\langle \bar dd\rangle=(-0.24 \pm 0.01)^3\,\mbox{GeV}^3$, $\langle \bar ss\rangle = (0.8 \pm 0.1)\, \langle \bar uu\rangle$ $\,\mbox{GeV}^3$ \cite{Ioffe:2005ym}, $m_0^{2} = 0.8 \pm 0.1 \,\mbox{GeV}^2$ \cite{Ioffe:2005ym},  and $\langle g_s^2G^2\rangle = 0.48 \pm 0.14~ \mbox{GeV}^4$~\cite{Narison:2018nbv}. Furthermore, the residues of the $P_{c(s)}$ states are required, as they are borrowed from Ref.~\cite{Wang:2015ixb}.  In numerical calculations, the parameters $m_u$, $m_d$, and $m_s^2$ are set to zero, since their contributions are found to be substantially small. However, it is important to note that terms proportional to $m_s$ are taken into account. To carry out further calculations, it is necessary to use the photon DAs and their explicit form, as well as the required numerical quantities, as described in Ref.~\cite{Ball:2002ps}. 

\begin{widetext}
%\newpage
 
 \begin{figure}[htp]
\centering
\subfloat[]{\includegraphics[width=0.45\textwidth]{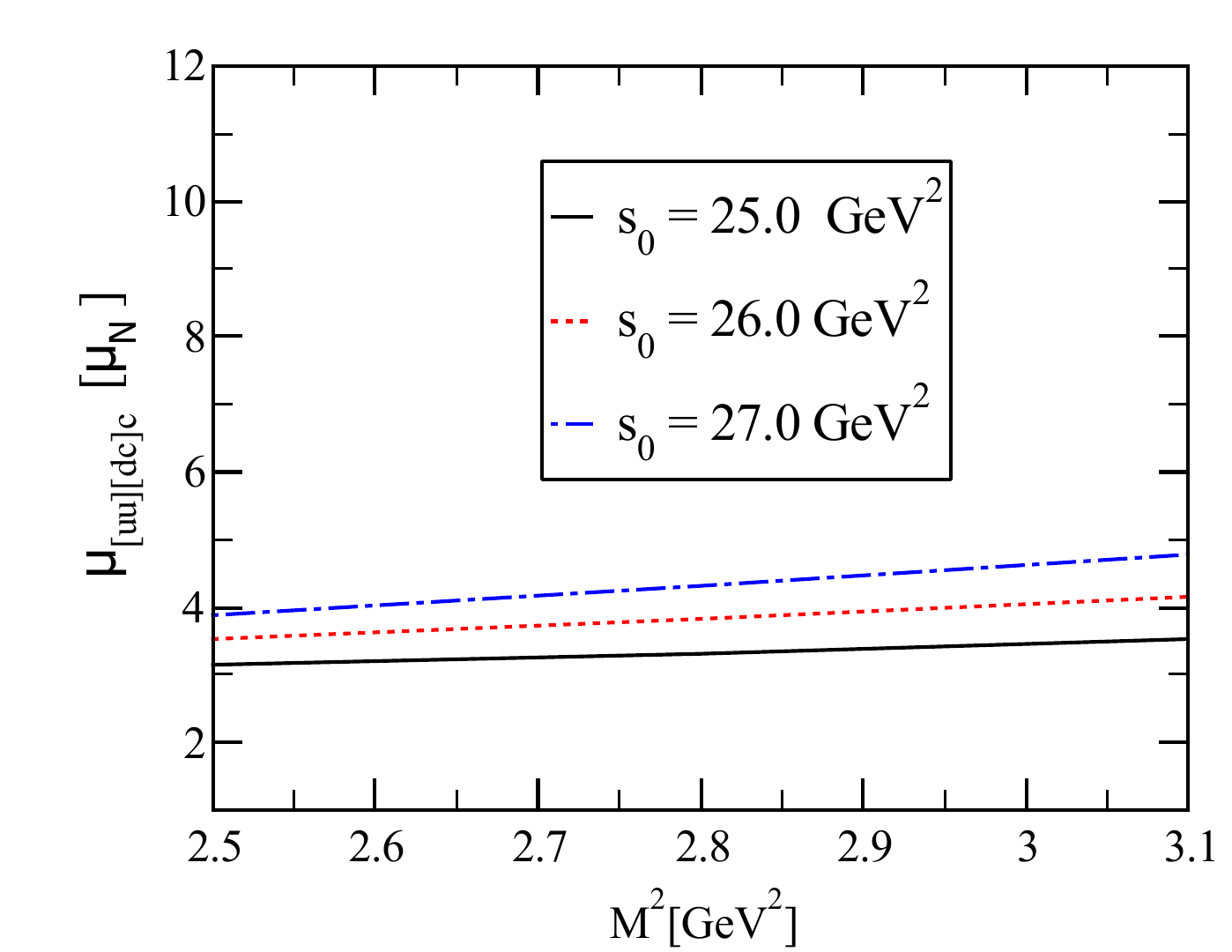}} ~~~~~~
\subfloat[]{\includegraphics[width=0.45\textwidth]{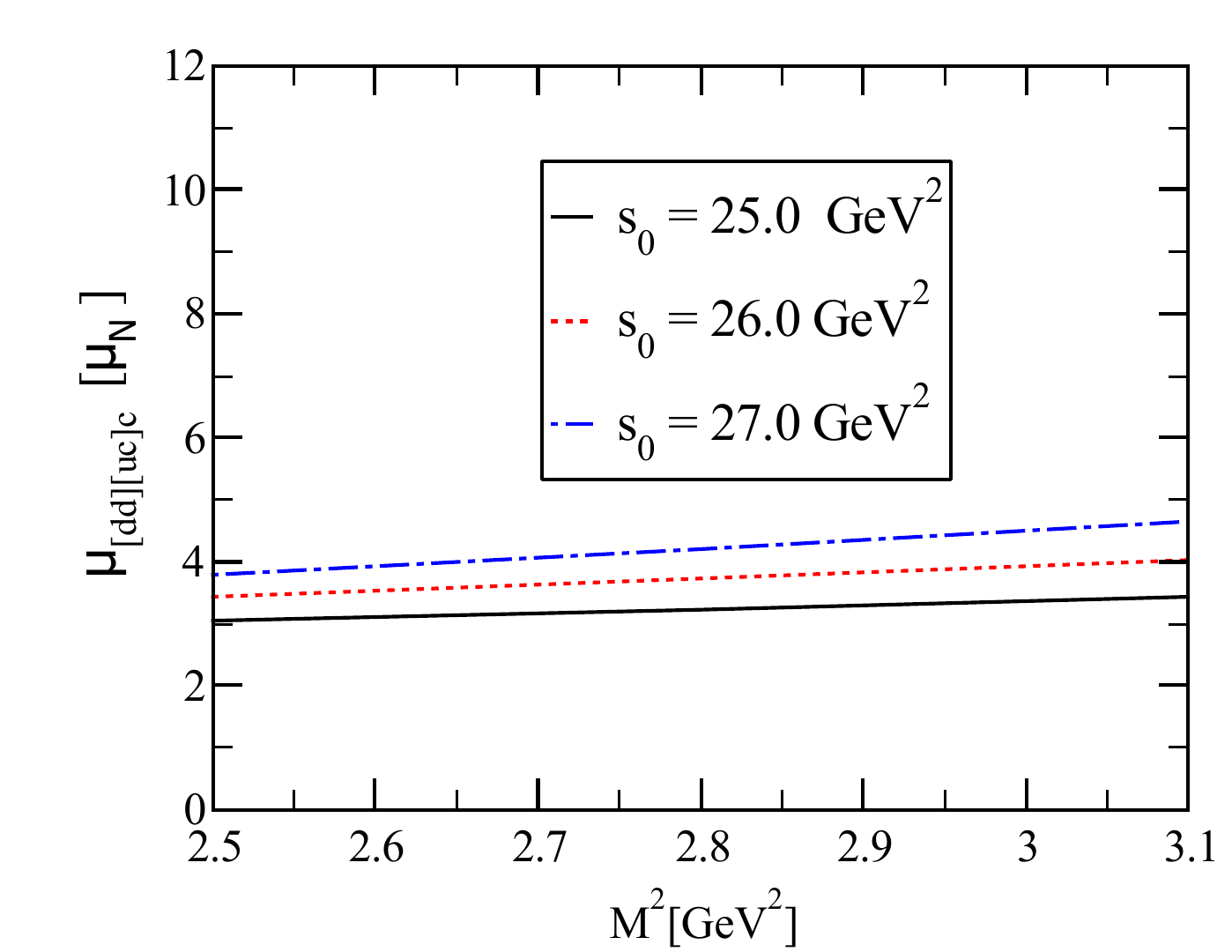}} \\
\subfloat[]{\includegraphics[width=0.45\textwidth]{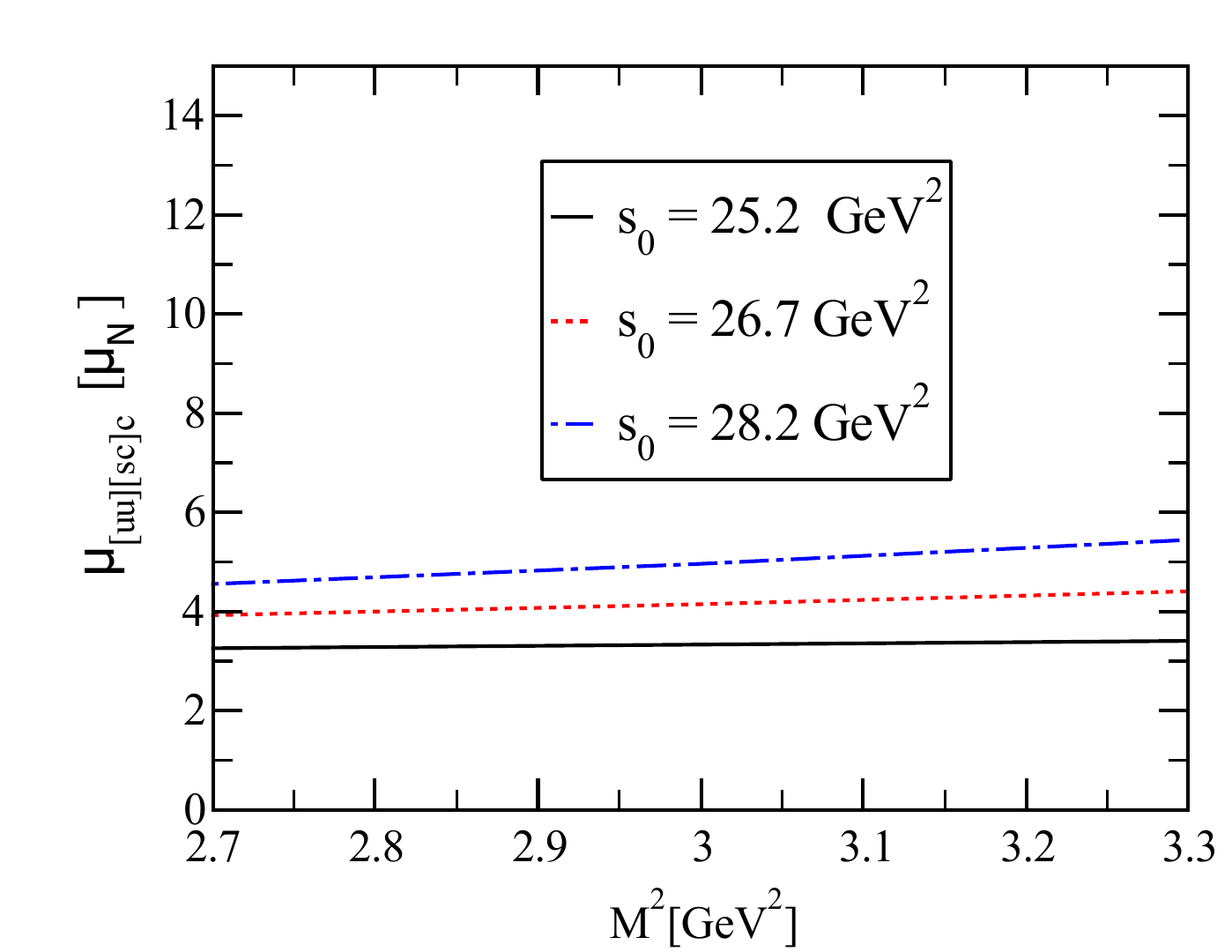}} ~~~~~~
\subfloat[]{\includegraphics[width=0.45\textwidth]{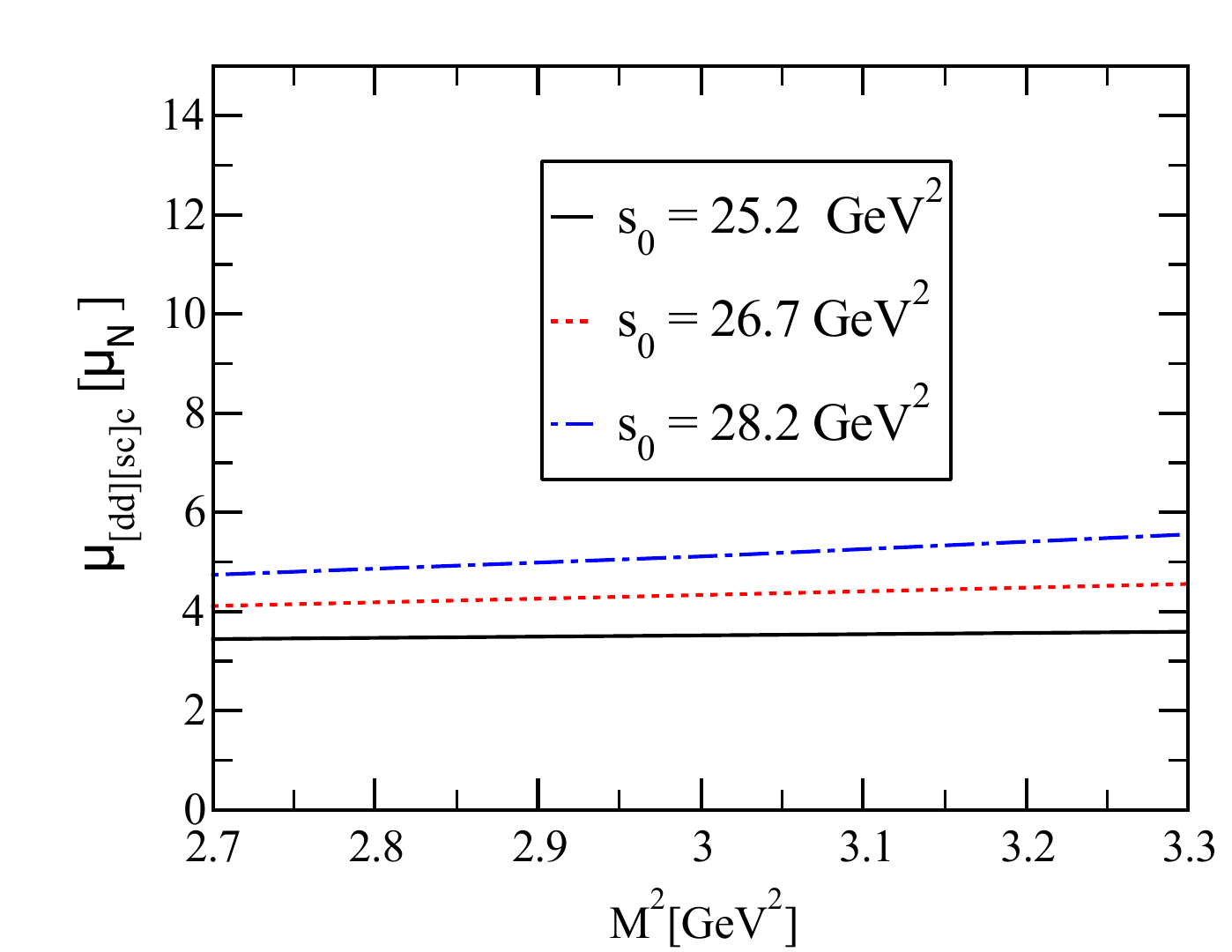}} \\
\subfloat[]{\includegraphics[width=0.45\textwidth]{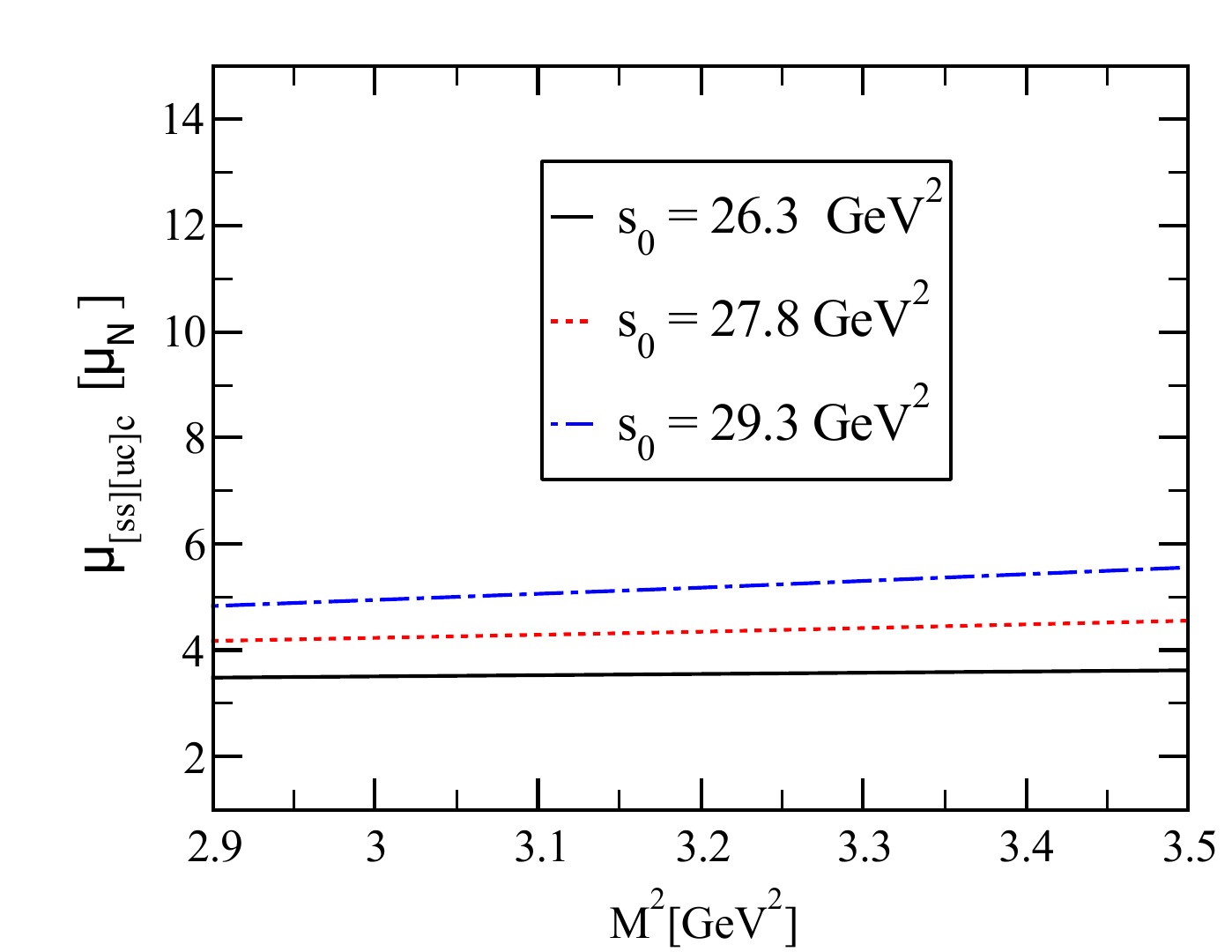}} ~~~~~~
\subfloat[]{\includegraphics[width=0.45\textwidth]{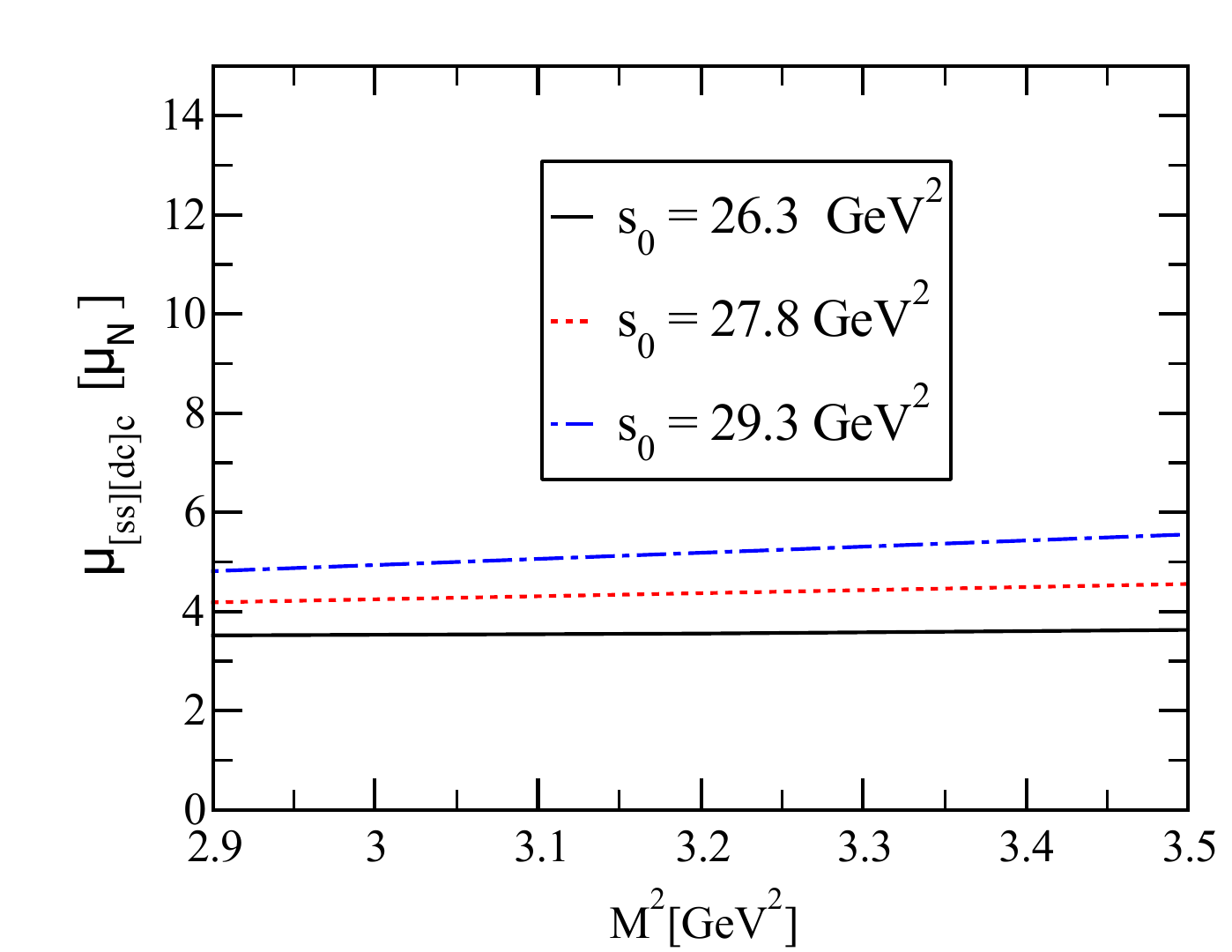}} 
 \caption{The magnetic dipole moments of the $P_{c(s)}$ states versus $\mathrm{M^2}$ for the $J_\mu^1$ current. }
 \label{Msqfig}
  \end{figure}

  \end{widetext}

\begin{widetext}
%\newpage
 
 \begin{figure}[htp]
\centering
\subfloat[]{\includegraphics[width=0.45\textwidth]{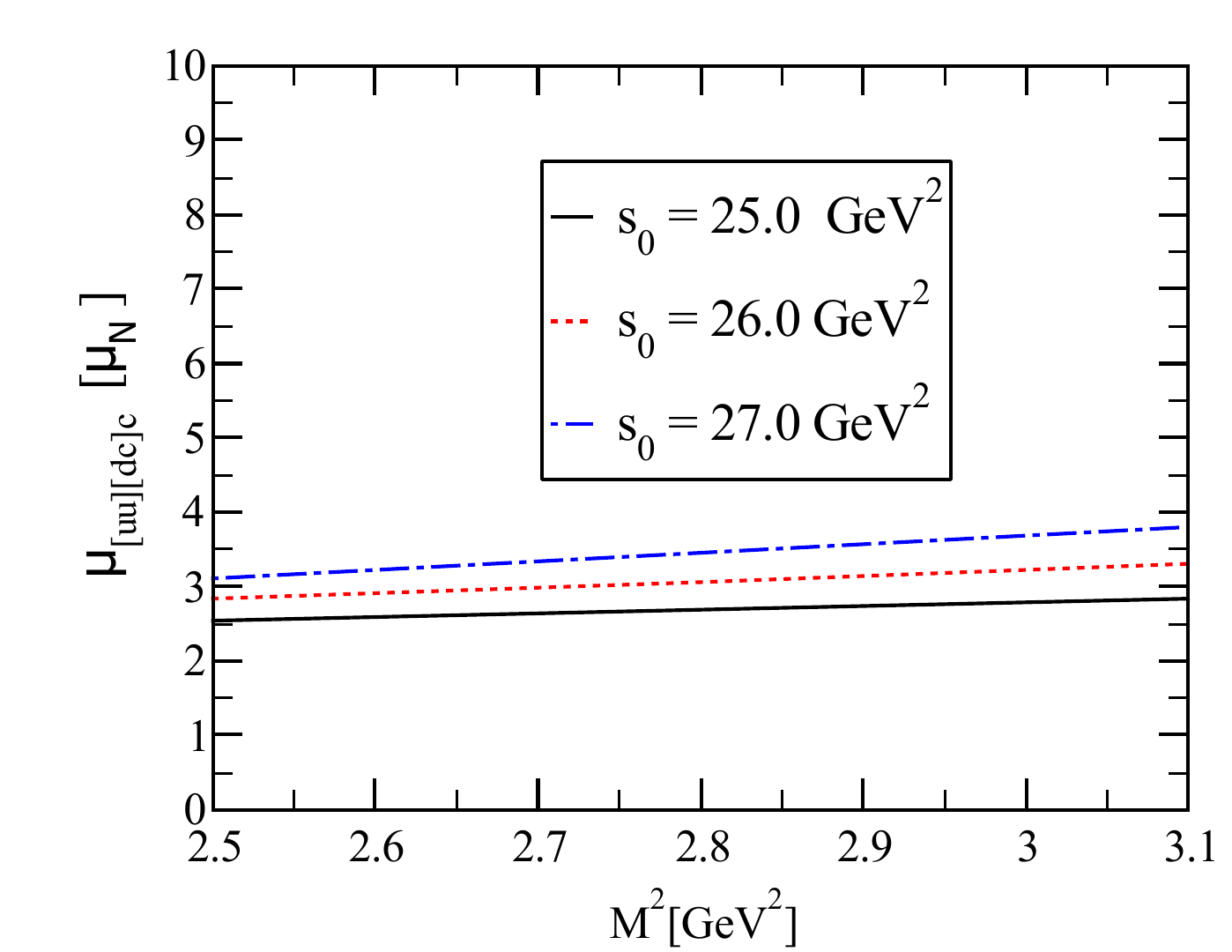}} ~~~~~~
\subfloat[]{\includegraphics[width=0.45\textwidth]{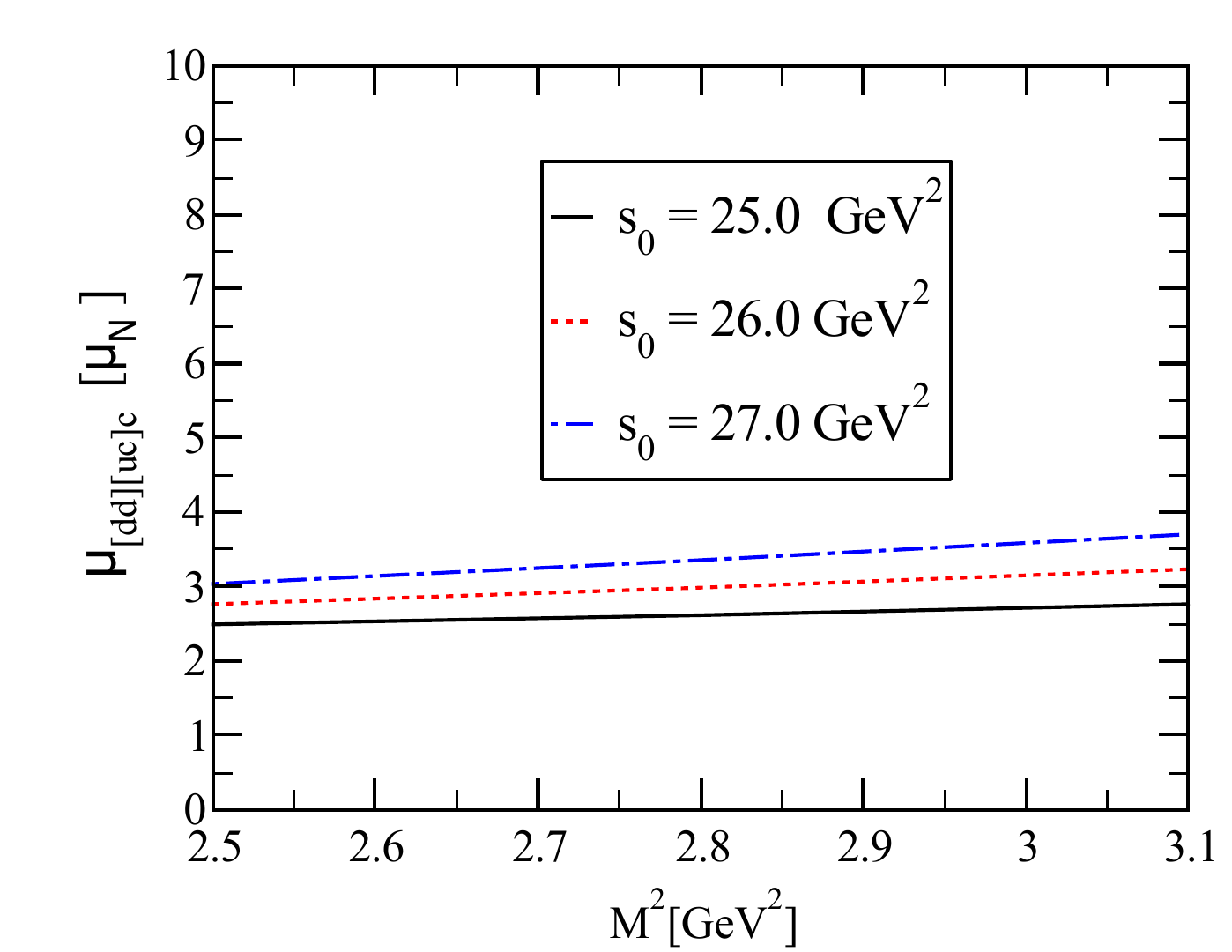}} \\
\subfloat[]{\includegraphics[width=0.45\textwidth]{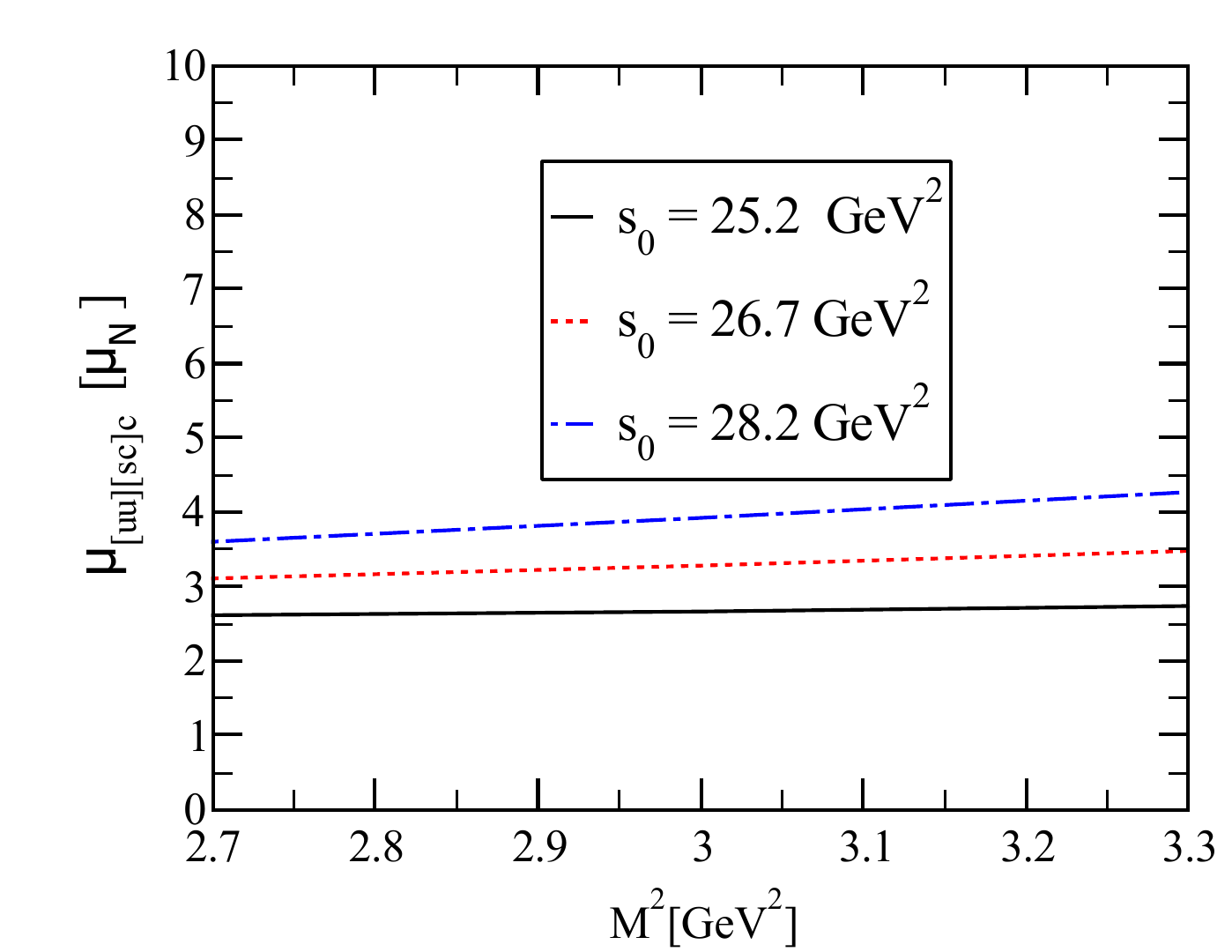}} ~~~~~~
\subfloat[]{\includegraphics[width=0.45\textwidth]{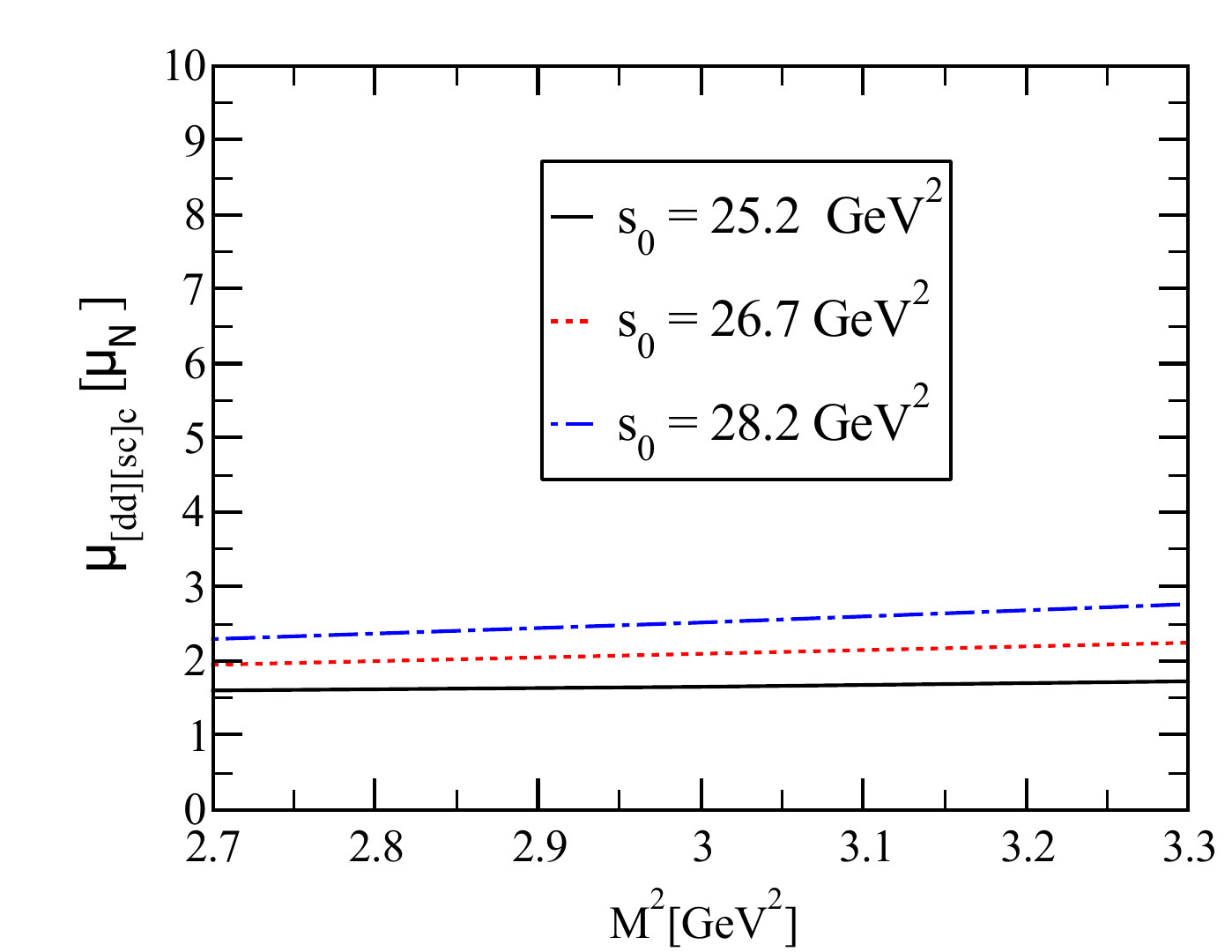}} \\
\subfloat[]{\includegraphics[width=0.45\textwidth]{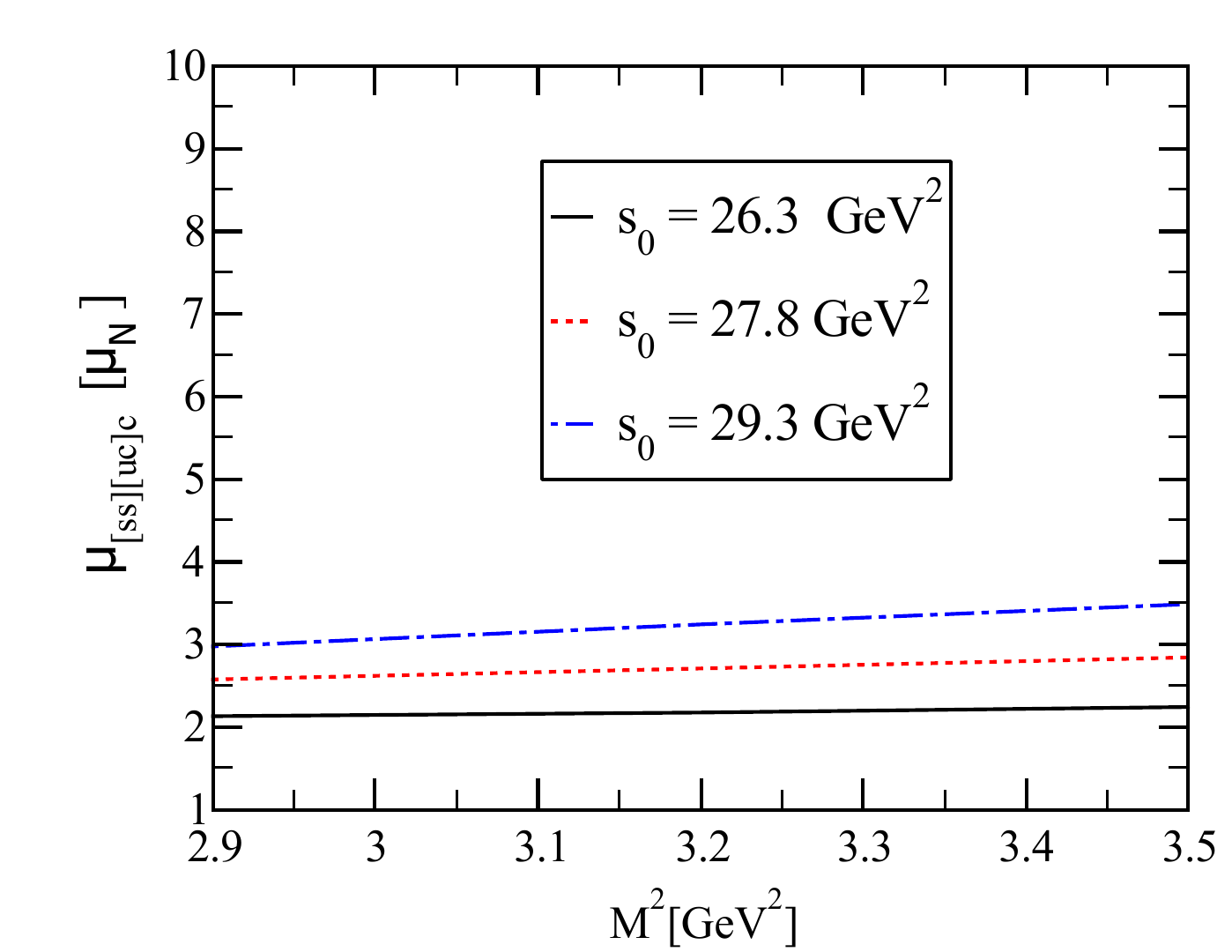}} ~~~~~~
\subfloat[]{\includegraphics[width=0.45\textwidth]{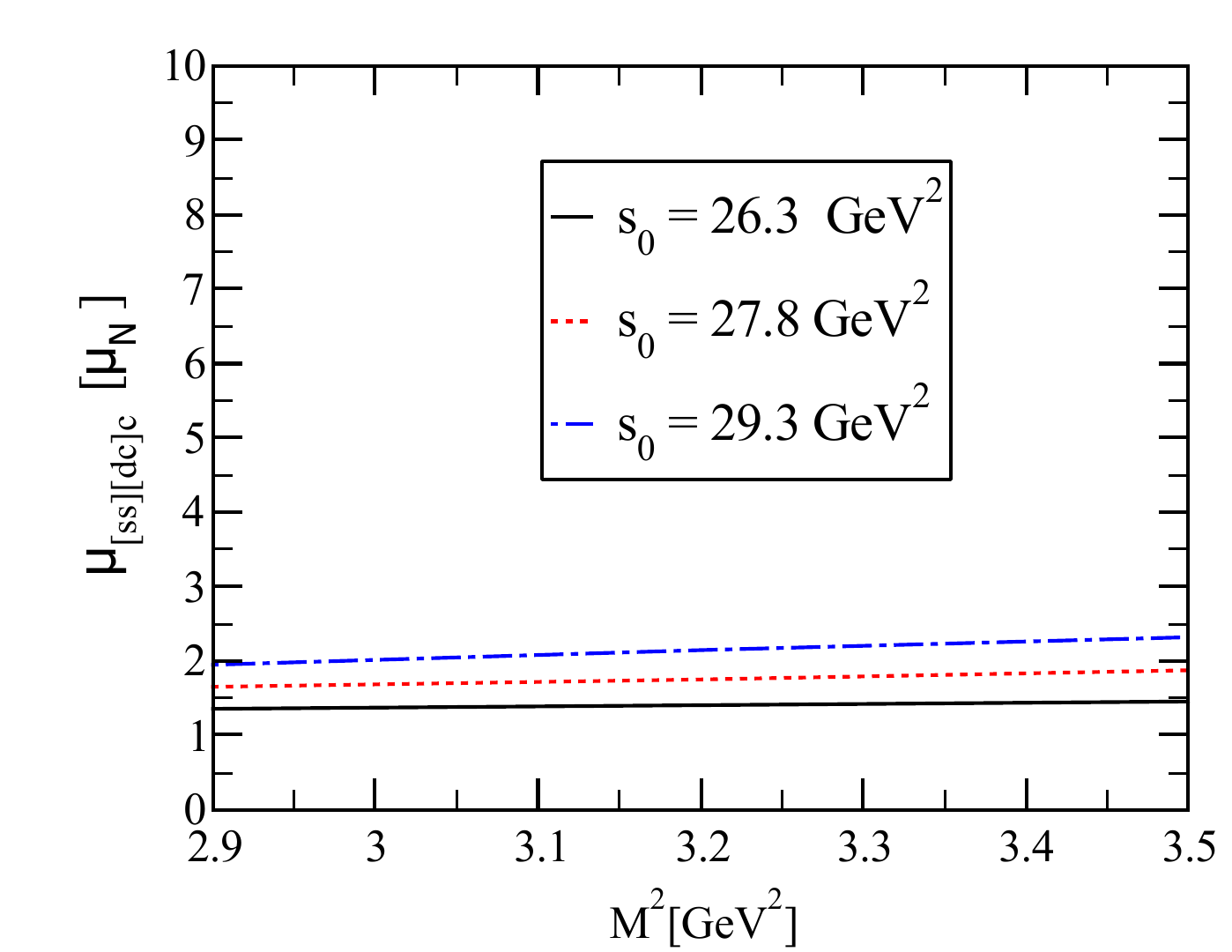}} 
 \caption{The magnetic dipole moments of the $P_{c(s)}$ states versus $\mathrm{M^2}$ for the $J_\mu^2$ current. }
 \label{Msqfig1}
  \end{figure}

  \end{widetext}

In light of the preceding numerical input variables, two supplementary parameters must be incorporated into the execution of this numerical analysis. The first of these is the continuum threshold parameter, designated as $\mathrm{s_0}$. The second is the Borel mass parameter, denoted by $\mathrm{M^2}$. In an ideal scenario, the numerical analysis should be performed independently of the parameters mentioned above. However, this approach is neither pragmatic nor realistic.  The establishment of a region of analysis is imperative to ensure that the impact of parameter variation on the numerical results is deemed negligible. 
The parameter $\mathrm{s_0}$ is not arbitrary; rather, it establishes a scale. Subsequent to this scale, continuum and higher states begin to exert an influence on the correlation function. Despite the existence of various methodologies described in the extant literature for determining the working region of this parameter, it is generally observed that this parameter varies within the range $(m_{P_{c(s)}} +0.5)^2$ GeV$^2$ $\leq$ $\mathrm{s_0}$ $\leq$ $(m_{P_{c(s)}}  +0.8)^2$ GeV$^2$.  Therefore, the approach that this parameter varies in this working interval is preferred.
  The working region for the $\mathrm{M^2}$, i.e. the range in which the variation in our numerical predictions for this variable is relatively small, is constrained by the applied approaches. The aforementioned constraints are referred to as pole contribution (PC) and convergence of OPE (CVG). Following the prevailing methodology, the CVG must demonstrate adequate smallness to assure convergence of the OPE.  Conversely, the PC is assumed to be adequately large to improve the efficiency of the single-pole scheme. The following formulas have been utilized to quantify them:
\begin{align}
 \mathrm{PC} &=\frac{\rho_i (\mathrm{M^2},\mathrm{s_0})}{\rho_i (\mathrm{M^2},\infty)} \geq 30 \%, ~~~~~~%\\
 %\nonumber\\
 \mathrm{CVG}=\frac{\rho_i^{\mathrm{DimN}} (\mathrm{M^2},\mathrm{s_0})}{\rho_i (\mathrm{M^2},\mathrm{s_0})} \leq 5 \%,
 \end{align}
 where $\rho_i^{\mbox{DimN}} (\rm{M^2},\rm{s_0})$ represent the highest dimensional terms in the $\rho_i (\rm{M^2},\rm{s_0})$. %DimN $\geq (8+9+10)$.  
 As demonstrated in the analytical expressions provided in the appendix, our analysis incorporates combinations of condensates with varying compositions, such as $\langle g_s^2G^2\rangle  \langle \bar q q\rangle^2$, $m_0^2$ $\langle g_s^2G^2\rangle  \langle \bar q q\rangle$, $\langle g_s^2G^2\rangle  \langle \bar q q\rangle$, $ \langle \bar q q\rangle^2$,  $\langle g_s^2G^2\rangle$, and $\langle \bar q q\rangle$.  
   In our analysis, the highest dimensional terms are dimension 8 ($m_0^2$ $\langle \bar q q\rangle^2$), 9 ($m_0^2$ $\langle g_s^2G^2\rangle  \langle \bar q q\rangle$), and 10 ($\langle g_s^2G^2\rangle  \langle \bar q q\rangle^2$). Consequently, the CVG analysis has been performed by considering the DimN expression in the form of $\rm{Dim(8+9+10)}$.  
 As a result of our analysis, taking into account the results given in Table~\ref{parameter}, it can be seen that the working regions determined for the parameters $\mathrm{s_0}$ and $\mathrm{M^2}$ are successful in satisfying the constraints of the method. To support the consistency of these study regions, the magnetic dipole moments at different values of $\mathrm{s_0}$ are plotted versus $\mathrm{M^2}$ in Figs.~\ref{Msqfig} and \ref{Msqfig1}. As demonstrated in these figures, the variation of the magnetic dipole moment resulting from these parameters is within an acceptable limit. It is also worth noting that the results may be affected by uncertainty due to the presence of residual dependencies.
  \begin{widetext}

  \begin{table}[t]
	\addtolength{\tabcolsep}{10pt}
	\caption{Working regions of $\rm{s_0}$
and $\rm{M^2}$ together with the CVG and PC for the magnetic moments of the $P_{c(s)}$ states, where the "Pert"  and "NPert"  stand for the contributions from the perturbative and non-perturbative terms, respectively.}
	\label{parameter}
		\begin{center}
		%\begin{ruledtabular}
\begin{tabular}{l|ccccccc}
	   \hline\hline
	   \\
  Current& State&  $\mathrm{s_0}$\mbox{(GeV$^2$)}&$\mathrm{M^2}$\mbox{(GeV$^2$)}&\mbox{CVG}$(\%)$&\mbox{PC}$(\%)$&\mbox{Pert}$(\%)$& NPert$(\%)$ \\
   \\
\hline\hline
&$[uu] [dc] \bar c$ &$ [25.0, 27.0]$         &  $ [2.5, 3.1] $& $\ll 1$&[56.45, 35.38]& $(88-93)$& $(7-12)$\\
&$[dd] [uc] \bar c$ &$ [25.0, 27.0]$         &  $ [2.5, 3.1] $& $\ll 1$&[56.36, 35.27]& $(88-93)$& $(7-12)$\\
%\\
&$[uu] [sc] \bar c$ &$ [25.2, 28.2]$         &  $ [2.7, 3.3] $& $\ll 1$&[56.15, 34.60] & $(89-94)$& $(6-11)$\\
%\\
$J_\mu^1$&$[dd] [sc] \bar c$ &$ [25.2, 28.2]$         &  $ [2.7, 3.3] $& $\ll 1$&[56.12, 35.78]& $(87-92)$ & $(8-13)$\\
%\\
&$[ss] [uc] \bar c$ &$ [26.3, 29.3]$         &  $ [2.9, 3.5] $& $\ll 1$&[55.08, 34.90]& $(89-94)$ & $(6-11)$\\
%\\
&$[ss] [dc] \bar c$ &$ [26.3, 29.3]$         &  $ [2.9, 3.5] $& $\ll 1$&[56.30, 36.05]& $(87-92)$ & $(8-13)$\\
%\\
	   \hline\hline
	   %%%%%%%%%%%%%%%%%%%%%%%%%%%%%%%%%%%%%%%%%%%%%%
	   	   \\
  Current& State&  $\mathrm{s_0}$\mbox{(GeV$^2$)}&$\mathrm{M^2}$\mbox{(GeV$^2$)}&\mbox{CVG}$(\%)$&\mbox{PC}$(\%)$&\mbox{Pert}$(\%)$& NPert$(\%)$ \\
   \\
\hline\hline
&$[uu] [dc] \bar c$ &$ [25.0, 27.0]$         &  $ [2.5, 3.1] $& $< 1.5$&[57.45, 35.38]& $(81-86)$ & $(14-19)$\\
&$[dd] [uc] \bar c$ &$ [25.0, 27.0]$         &  $ [2.5, 3.1] $& $< 1.5$&[57.36, 35.27]& $(81-86)$ & $(14-19)$\\
%\\
&$[uu] [sc] \bar c$ &$ [25.2, 28.2]$         &  $ [2.7, 3.3] $& $< 1.5$&[57.23, 35.47]& $(80-85)$ & $(15-20)$\\
%\\
$J_\mu^2$&$[dd] [sc] \bar c$ &$ [25.2, 28.2]$         &  $ [2.7, 3.3] $& $< 1.5$&[54.17, 35.24]& $(75-82)$ & $(18-25)$\\
%\\
&$[ss] [uc] \bar c$ &$ [26.3, 29.3]$         &  $ [2.9, 3.5] $& $< 1.5$&[53.97, 34.14]& $(80-85)$ & $(15-20)$\\
%\\
&$[ss] [dc] \bar c$ &$ [26.3, 29.3]$         &  $ [2.9, 3.5] $& $< 1.5$&[53.64, 33.17]& $(75-82)$& $(18-25)$\\
%\\
	   \hline\hline
\end{tabular}
\end{center}
%\end{ruledtabular}
\end{table}

\end{widetext}
 
 \begin{widetext}
 
  \begin{table}[htb!]
	\addtolength{\tabcolsep}{10pt}
	\caption{Predictions of the electromagnetic multipole moments of the $P_{c(s)}$ states.}
	\label{parameter1}
		\begin{center}
		%\begin{ruledtabular}
\begin{tabular}{l|ccccc}
	   \hline\hline
	  \\
   Current& Pentaquarks& $\mu\,(\mu_N)$& $\mathcal Q$\,($10^{-2}$\,fm$^2$) & $\mathcal O$\,($10^{-3}$\,fm$^3$)\\
   \\
\hline\hline
&$[uu] [dc] \bar c$& $ 3.95 \pm 0.82 $ &$ 0.31 \pm 0.02$         &  $- 0.56 \pm 0.07 $
\\
&$[dd] [uc] \bar c$& $ 3.86 \pm 0.80 $ &$ 1.79 \pm 0.23$         &  $- 0.81 \pm 0.07 $
\\
&$[uu] [sc] \bar c$& $ 4.33 \pm 1.09 $ &$ 0.32 \pm 0.02$         &  $- 0.58 \pm 0.08 $\\
%\\
$J_\mu^1$&$[dd] [sc] \bar c$& $ 4.50 \pm 1.04$ &$3.48 \pm 0.60 $       &  $ -1.09 \pm 0.08 $\\
%\\
&$[ss] [uc] \bar c$& $ 4.53 \pm 1.03$ &$ 1.83 \pm 0.28$         &  $ -0.80 \pm 0.08$\\
%\\
&$[ss] [dc] \bar c$& $4.48 \pm 0.98$ &$3.42 \pm 0.56$         &  $ -1.04 \pm 0.08 $\\
%\\
	   \hline\hline
	   %%%%%%%%%%%%%%%%%%%%%%%%%%%%%%%%%%%%%%%%%%%%%%%%%%%%%%%
	   	   \\
    Current& Pentaquarks& $\mu\,(\mu_N)$& $\mathcal Q$\,($10^{-2}$\,fm$^2$) & $\mathcal O$\,($10^{-3}$\,fm$^3$)\\
   \\
\hline\hline
&$[uu] [dc] \bar c$& $ 3.17 \pm 0.82 $ &$ -1.43 \pm 0.26$         &  $- 0.20 \pm 0.04 $\\
&$[dd] [uc] \bar c$& $ 3.09 \pm 0.61 $ &$ ~~1.51 \pm 0.19$         &  $- 0.62 \pm 0.05 $\\
&$[uu] [sc] \bar c$& $ 3.43 \pm 0.83 $ &$ -1.53 \pm 0.33$         &  $- 0.21 \pm 0.05 $\\
%\\
$J_\mu^2$&$[dd] [sc] \bar c$& $ 2.19 \pm 0.59$ &$~~4.85 \pm 0.88 $       &  $ -1.12 \pm 0.07 $\\
%\\
&$[ss] [uc] \bar c$& $ 2.81 \pm 0.67$ &$ ~~1.57 \pm 0.25$         &  $ -0.63 \pm 0.06$\\
%\\
&$[ss] [dc] \bar c$& $1.84 \pm 0.49$ &$~~4.80 \pm 0.82$         &  $ -1.06 \pm 0.06 $\\
%\\
	   \hline\hline
\end{tabular}
\end{center}
%\end{ruledtabular}
\end{table}

\end{widetext}

 All the necessary variables for the numerical evaluation are determined. The entire numerical predictions, together with all the inherent uncertainties concerning the input quantities, have been listed in Table~\ref{parameter1}.
The following conclusions can be delineated from the results obtained: 
 \begin{itemize}
 
 % \item The primary contributions to the electromagnetic multipole moments of the $P_{c(s)}$ states in QCD light-cone sum rules are derived from the perturbative component of the spectral density. The perturbative component can be represented schematically as: $( A \, e_c \pm B \, (2 e_{q_1}+e_{q_2}))$. In the case of the magnetic dipole moment, the contribution of the $B \, (2 e_{q_1}+e_{q_2})$ term is found to be quite small. In contrast, the contribution of this term to the electric quadrupole and magnetic octupole moments is found to be larger, even large enough to determine the sign of the electric quadrupole moments in some $P_{c(s)}$ states.  
 
 \item In the QCD light-cone sum rules framework, the primary contributions to the electromagnetic multipole moments of the $P_{c(s)}$  states originate from the perturbative component of the spectral density. This component can be schematically expressed as  $( A \, e_c \pm B \, (2 e_{q_1}+e_{q_2}))$. For the magnetic dipole moment, the contribution from the $ B \, (2 e_{q_1}+e_{q_2})$  term is found to be negligible. However, for the electric quadrupole and magnetic octupole moments, this term plays a more significant role, in some cases even determining the sign of the electric quadrupole moment in certain $P_{c(s)}$  states.

 \item A more thorough investigation of the magnetic dipole moment requires an examination of the contributions of light quarks and the charm quark. The findings of this comprehensive analysis are presented in Table \ref{parameter2}, where the central values of all input parameters are used. The contributions of both light quarks and the charm quark differ significantly depending on the choice of interpolating current.     In the case of the $J_\mu^1$ interpolating current, the contributions of light quarks to the magnetic dipole moment are quite small. In contrast, in the case of the $J_\mu^2$ interpolating current, these contributions become larger than in the $J_\mu^1$ interpolating current. The analysis of quark contributions to the magnetic dipole moments reveals that the charm quark dominates the magnetic properties of pentaquarks. 
 The light quarks (up, down, and strange) contribute less significantly to the magnetic dipole moments, but their contributions vary depending on the diquark configuration. This variation suggests that the light quark dynamics are sensitive to the specific arrangement of quarks within the pentaquark.
 %The dominance of the charm quark in the magnetic dipole moment suggests that the overall magnetic properties of these pentaquarks are primarily shaped by the charm quark. However, the contributions from light quarks cannot be ignored, as they are influenced by the internal diquark configurations.
  
  \item  The contribution of charm- and light quarks to the magnetic dipole moment have been observed to exhibit an inverse relationship, except for the $[u u][d c] \bar c$, $[dd][u c] \bar c$, and $[uu][sc] \bar c$ state, which is obtained in the case of $J_\mu^2$. The signs of the magnetic dipole moments demonstrate the interaction of the spin degrees of freedom of the quarks. %The opposing signs of the charm- and light quarks magnetic dipole moments indicate that their spins are anti-aligned in the $P_{c(s)}$ state.
  The opposite signs of the magnetic dipole moments for charm and light quarks suggest that their spins are anti-aligned within the pentaquark. 
 
  \item The electric quadrupole and magnetic octupole moments of the related pentaquarks are also calculated. The obtained results are given in the Table \ref{parameter1}. It may be posited that the values of the electric quadrupole and magnetic octupole moments are substantially smaller than those of the magnetic dipole moments.  This finding indicates the presence of non-zero values for the electric quadrupole and magnetic octupole moments of the observed pentaquarks, suggesting a non-spherical charge distribution. 
  The sign of the electric quadrupole moment reveals the overall shape of the charge distribution in the baryon states: Positive quadrupole moment-indicates a prolate deformation—charge is stretched along the quantization axis. Negative quadrupole moment-indicates an oblate deformation—charge is concentrated in the transverse plane.  The magnetic octupole moment gives information about the spatial distribution of the magnetic current inside the baryon: Positive octupole moment- suggests that magnetic currents are aligned or concentrated along a specific axis. Negative octupole moment-indicates an opposite orientation of the magnetic current distribution. 
  It can be posited that the signs of the higher multipole moments can offer insights into the deformation of the corresponding hadron and its direction. If the signs of the higher multipole moments are the same, it indicates that both the geometric deformation and the charge distribution are aligned in the same direction. Conversely, if the signs differ, this implies that the spatial deformation and the charge distribution are oriented oppositely. In the event of a negative prediction, the shape of the hadron is classified as an oblate, while positive predictions are indicative of prolate shapes. The sign of the electric quadrupole moments is predicted to be positive, except for the $[u u][d c] \bar c$, and $[uu][sc]\bar c$ state, which are obtained in the case of $J_\mu^2$ and corresponds to the prolate charge distribution.  In contrast, the sign of the magnetic octupole moments is obtained as negative. This finding suggests that the charge distribution and geometrical shape of these states are opposite. However, in the case of the $[u u][d c] \bar c$, and $[uu][sc] \bar c$ state, which are obtained in the case of $J_\mu^2$, it is observed that both the electric quadrupole and magnetic octupole moments have the same sign and geometric shape as the charge distribution (oblate). As can be seen from the obtained results, different diquark configurations affect not only the geometric shape of the corresponding hadrons but also their internal charge distributions.

  \item To guarantee the comprehensiveness of the analysis, the contribution of light quarks and the c-quark to the electric quadrupole and magnetic octupole moments are also investigated. The predicted values of these parameters can be found in Tables \ref{parameter3} and \ref{parameter4}. In the quark contribution analysis, a similar result to that observed in the magnetic dipole moment output is evident. It can be seen that the contributions of both light quarks and c-quarks are quite different in the two interpolating currents used. The contributions of the light quarks in the second interpolating current have been obtained to be approximately twice those of the first interpolating current. As can be seen from the results here, the different diquark structures significantly influence the outcomes.

  \item Given that U-spin symmetry-breaking effects have been incorporated through a nonzero strange-quark mass and strange-quark condensate, as reflected in the ratios of electromagnetic multipole moments $(
\frac{[uu][dc]\bar c}{[uu][sc]\bar c}$, and   $\frac{[dd][uc]\bar c}{[ss][uc]\bar c}$), the resulting predictions indicate a U-spin violation of at most $15\%$. The typical U-spin symmetry breaking effects are anticipated to reach, at most, the level of approximately 30$\%$. 
A detailed examination of these symmetry violation effects reveals that the maximum violation reaches approximately $15\%$ for the $J_\mu^1$ interpolating current, while for the $J_\mu^2$ interpolating current, the violation is slightly lower, around $10\%$. This difference may stem from the varying sensitivity of the interpolating currents to strange-quark contributions, as each current emphasizes different quark-gluon configurations and coupling structures. These findings indicate a moderate but noticeable violation of U-spin symmetry, consistent with theoretical expectations.
Moreover, the analysis shows that the construction of interpolating currents with different internal structures alters the relative contributions of strange-quark terms in the operator product expansion. These results also demonstrate that the choice of diquark configurations plays a significant role in modulating the extent and manifestation of U-spin symmetry-breaking effects. The diquark structure affects the flavor composition and internal correlations of the baryon, which in turn influences the sensitivity to strange-quark dynamics. 
Taken together, our findings confirm that U-spin symmetry is moderately broken at a level below 15$\%$, but that its breaking is not negligible. These effects should be considered in any precise determination of the internal electromagnetic structure of hadrons.

\item   The magnetic dipole moments obtained using different interpolating currents are related as follows: 
\[
\mu_{J_\mu^1} \sim \lambda \times \mu_{J_\mu^2}
\]
where \( \lambda \) varies for different pentaquark states as shown in the table below:

\begin{table}[h]
    \centering
    \renewcommand{\arraystretch}{1.3} % Adjust row height for better readability
    \setlength{\tabcolsep}{10pt}      % Adjust column spacing
    \begin{tabular}{|c|c|}
        \hline\hline
        \textbf{Pentaquark State} & \textbf{\( \lambda \) (Scaling Factor)} \\
        \hline\hline
        \( [uu][dc]\bar{c} \) & \( 1.25 \) \\
        \( [dd][uc]\bar{c} \) & \( 1.25 \) \\
        \( [uu][sc]\bar{c} \) & \( 1.25 \) \\
        \( [dd][sc]\bar{c} \) & \( 2.05 \) \\
        \( [ss][uc]\bar{c} \) & \( 1.60 \) \\
        \( [ss][dc]\bar{c} \) & \( 2.40 \) \\
        \hline\hline
    \end{tabular}
    \caption{Scaling factors (\(\lambda\)) for different pentaquark states.}
    \label{tab:scaling_factors}
\end{table}
  As evidenced by the aforementioned equations, the different diquark configurations constructed for pentaquarks result in a substantial variation in the observed outcomes.
  
  \item   As demonstrated in Table ~\ref{parameter2}, the application of different interpolating currents to the $P_{c(s)}$ states—despite having identical quark content and quantum numbers—leads to substantial discrepancies in the obtained magnetic dipole moments. These results may suggest the existence of multiple $P_{c(s)}$ states with the same quantum numbers, yet differing in their electromagnetic properties due to variations in their internal quark configurations.
  While these states are nearly degenerate in mass~\cite{Wang:2015ixb}, their electromagnetic properties appear to be highly sensitive to the underlying diquark configurations. This raises important questions about the common assumption that changing the basis of hadrons does not affect physical observables. In the context of electromagnetic properties, a change in the hadronic basis may correspond to a modification in the internal structure, leading to significant variations in the predicted results.
  Previous studies~\cite{Ozdem:2024txt, Ozdem:2024dbq, Azizi:2023gzv, Ozdem:2024rqx, Ozdem:2022iqk, Ozdem:2024rch} have explored the electromagnetic properties of tetraquark and pentaquark states using various interpolating currents, revealing substantial deviations in the magnetic dipole moments obtained from different diquark-antidiquark and diquark-diquark-antiquark configurations. Consequently, the choice of interpolating currents—or equivalently, the isospin and charge basis of the studied hadrons—may significantly impact the electromagnetic moments.

  \item The mass of the $[u u][d c] \bar c$ state is very close to the mass of the $P_c(4440)$ pentaquark state within errors. One can therefore consider the possibility that these interpolating currents couple to this $P_c(4440)$ pentaquark state.  Comparing the results obtained for state $P_c(4440)$ with the results available in the literature can provide a useful consistent perspective.
  In Ref.~\cite{Li:2021ryu}, the magnetic dipole moment of the $P_{c}(4440)$ state has been examined within the quark model, interpreted as being in the molecular configuration with the quantum numbers $\mathrm{J^P =\frac{1}{2}^-}$. %In the analysis conducted, this pentaquark was interpreted as being in the molecular configuration.  
  The resultant value is referred to as $\mu_{P_{c}(4440)} =- 0.979~\mu_N$. 
  In Ref.~\cite{ Ozdem:2024jty}, the magnetic dipole moment of the $P_{c}(4440)$ state has been studied utilizing the QCD light-cone sum rules for meson-baryon pentaquark picture with $\mathrm{J^P =\frac{3}{2}^-}$ quantum numbers. The resultant value is referred to as  $\mu_{P_{c}(4440)} =  0.73^{+0.26}_{-0.24}~\mu_N$. The observed differences in the magnetic dipole moments may stem from the distinct internal dynamics of these configurations. In the diquark-diquark-antiquark picture, the clustering of diquarks may lead to a different spin alignment compared to the meson-baryon interpretation, where the spatial separation between the meson and baryon components could modify the magnetic response. The differences in electromagnetic properties between these configurations (diquark-diquark-antiquark and meson-baryon molecular states) could be used to distinguish between them in future experimental studies.  The findings from the numerical analyses conducted within the present study suggest that the magnetic dipole moments of $P_{c}(4440)$ state may provide crucial insights into their fundamental structures. These results, in turn, may facilitate a more refined interpretation of the distinction between their respective spin-parity quantum numbers. Future studies should explore these differences to better understand the underlying structure of these exotic states.
  
 \item Determining the magnetic dipole moments of $P_{c(s)}$ pentaquarks via spin precession experiments presents significant challenges owing to their relatively short lifetimes. This difficulty stems from the need for stable, long-lived systems to allow for accurate measurements of spin dynamics. As such, direct observation is not feasible. Instead, the magnetic dipole moment of these $P_{c(s)}$ states can only be determined indirectly through a three-step procedure. First, the $P_{c(s)}$ is generated. Then, it emits a low-energy photon that acts as an external magnetic field. Finally, the $P_{c(s)}$ undergoes decay.  The magnetic dipole moment of the $\Delta$ resonance was obtained using this scheme through the $\gamma N \rightarrow \Delta \rightarrow \Delta \gamma \rightarrow \pi N \gamma$ process~\cite{Pascalutsa:2004je, Pascalutsa:2005vq, Pascalutsa:2007wb, Kotulla:2002cg, Drechsel:2001qu, Machavariani:1999fr, Drechsel:2000um, Chiang:2004pw, Machavariani:2005vn}. By analogy, a comparable process, $\gamma^{(*)}N $ $ \rightarrow $ $P_{c} $ $\rightarrow P_{c} \gamma$ $ \rightarrow $ $ J /\psi N \gamma$ or $\gamma^{(*)}\Lambda $ $ \rightarrow $ $P_{cs} $ $\rightarrow P_{cs} \gamma$ $ \rightarrow $ $ J/\psi \Lambda \gamma$, could be utilized to extract the magnetic dipole moment of the $P_{c(s)}$ pentaquarks. Future experiments at facilities like LHCb, Belle II, and PANDA could focus on measuring the angular distributions and polarization observables in pentaquark decays, which are sensitive to electromagnetic multipole moments. 
In parallel, lattice QCD calculations, which have been successfully employed to compute the magnetic dipole moments of baryons containing two charm quarks~\cite{Can:2013zpa, Can:2013tna}, are expected to provide crucial theoretical predictions. These methods, in combination with forthcoming experimental data, will help refine our understanding of the electromagnetic properties of exotic hadrons.

\end{itemize}

\section{Conclusions}\label{summary}

 In the present study, we investigate the magnetic dipole, electric quadrupole, and magnetic octupole moments of the $[u u][d c] \bar c$, $[dd][u c] \bar c$, $[u u][s c] \bar c$, $[dd] [s c] \bar c$, $[s s][u c] \bar c$, and $[s s][d c] \bar c$ states in the context of the QCD light-cone sum rules. In the course of examining these properties, two distinct diquark-diquark-antiquark forms of the interpolating currents are employed, given that these pentaquark candidates possess quantum numbers $\mathrm{J^P =\frac{3}{2}^-}$.  From the numerical results, we observe that there are significant discrepancies in the predicted magnetic dipole moments depending on the chosen diquark-diquark-antiquark structure. %This finding suggests the potential for multiple pentaquarks, each with identical quantum numbers and quark constituents but possessing distinct magnetic dipole moments. The resulting numerical values suggest that the magnetic dipole moments of hidden-charm pentaquark states may offer insights into their internal structure, which could potentially be utilized to ascertain their quark-gluon composition and quantum numbers. %It appears that in the context of future experimental searches for the family of hidden-charm pentaquark states, a rigorous examination of the electromagnetic properties of these states may yield significant insights. 
The results of this study provide valuable insights into the electromagnetic properties of hidden-charm pentaquarks, shedding light on their internal structure and quark-gluon dynamics. The significant differences in electromagnetic properties between different diquark configurations suggest that pentaquarks may exist in multiple states with similar quantum numbers but different internal structures. These findings have important implications for future experimental searches and theoretical studies of exotic hadrons. By combining theoretical predictions with experimental observations, we can move closer to a comprehensive understanding of the nature of pentaquarks and the fundamental forces that bind quarks together. Future theoretical studies should explore alternative pentaquark configurations, such as molecular models, for a better understanding of the range of possible electromagnetic properties. Additionally, lattice QCD calculations could provide more precise predictions for comparison with experimental data. 
The discrepancies between theoretical predictions and experimental observations (when available) serve as a testing ground for various QCD models. The ability of different models to reproduce the electromagnetic properties of pentaquarks could help refine our understanding of QCD dynamics in the non-perturbative regime.

%\newpage

\begin{widetext}
 
  \begin{table}[htb!]
	\addtolength{\tabcolsep}{10pt}
	\caption{The contribution of light and heavy quarks to the magnetic dipole moment of the $P_{c(s)}$ states ($\mu_N$).}
	\label{parameter2}
		\begin{center}
		%\begin{ruledtabular}
\begin{tabular}{l|ccccc}
	   \hline\hline
	%   \\
   Current&Pentaquarks& $\mu_{q}$&  $\mu_{c}$& $\mu_{total}$\\
   %\\
\hline\hline
%\\
&$[uu] [dc] \bar c$& $ -0.30 $    &  $4.25 $& $3.95$\\
&$[dd] [uc] \bar c$& $ -0.31 $    &  $4.17 $& $3.86$\\
&$[uu] [sc] \bar c$& $ -0.27 $    &  $4.60 $& $4.33$\\
%\\
$J_\mu^1$ &$[dd] [sc] \bar c$ &$-0.11 $   &  $ 4.61$&$ 4.50$\\
%\\
&$[ss] [uc] \bar c$& $-0.17$      &  $ 4.70$&$4.53$\\
%\\
&$[ss] [dc] \bar c$& $-0.21$        &  $ 4.69$&$4.48$\\
%\\
	   \hline\hline
	   %%%%%%%%%%%%%%%%%%%%%%%%
	%   \\
   Current&Pentaquarks& $\mu_{q}$&  $\mu_{c}$& $\mu_{total}$\\
   %\\
\hline\hline
&$[uu] [dc] \bar c$& $ ~~0.21 $   &  $2.96 $& $3.17$\\
&$[dd] [uc] \bar c$& $~~0.14$      &  $ 2.95$&$3.09$\\
%\\
&$[uu] [sc] \bar c$& $ ~~0.26 $   &  $3.17 $& $3.43$\\
%\\
$J_\mu^2$ &$[dd] [sc] \bar c$     &$-0.98 $   &  $ 3.17 $&$ 2.19$\\
%\\
&$[ss] [uc] \bar c$& $-0.41$      &  $ 3.22$&$2.81$\\
%\\
&$[ss] [dc] \bar c$& $-1.38$        &  $ 3.22$&$1.84$\\
%\\
	   \hline\hline
\end{tabular}
\end{center}
%\end{ruledtabular}
\end{table}

\end{widetext}
 \begin{widetext}
 
  \begin{table}[htb!]
	\addtolength{\tabcolsep}{10pt}
	\caption{The contribution of light and heavy quarks to the electric quadrupole moment of the $P_{c(s)}$ states ($10^{-2}$\,fm$^2$).}
	\label{parameter3}
		\begin{center}
		%\begin{ruledtabular}
\begin{tabular}{l|ccccc}
	   \hline\hline
	 %  \\
  Current& Pentaquarks& $\mathcal Q_{q}$ & $\mathcal Q_{c}$& $\mathcal Q_{total}$\\
  %\\
     \hline\hline
     &$[uu] [dc] \bar c$& $ -1.53 $     &  $1.84 $& $0.31$\\
%\\
&$[dd] [uc] \bar c$& $ ~~0.00 $     &  $1.79 $& $1.79$\\
&$[uu] [sc] \bar c$& $ -1.62 $     &  $1.94 $& $0.32$\\
%\\
$J_\mu^1$ &$[dd] [sc] \bar c$ &$~~1.54$       &  $ 1.94 $& 3.48\\
%\\
&$[ss] [uc] \bar c$& $ ~~0.00$         &  $ 1.83$& 1.83\\
%\\
&$[ss] [dc] \bar c$& $~~1.52$         &  $ 1.90$& 3.42\\
%\\
	   \hline\hline
	   	   %%%%%%%%%%%%%%%%%%%%%%%%
	%   	   \\
  Current& Pentaquarks& $\mathcal Q_{q}$&  $\mathcal Q_{c}$& $\mathcal Q_{total}$\\
  % \\
     \hline\hline
     &$[uu] [dc] \bar c$& $ -3.01 $ &  $1.58 $& $-1.43$\\
     &$[dd] [uc] \bar c$& $ ~~0.00$&  $ 1.51$& ~~1.51\\
%\\
&$[uu] [sc] \bar c$& $ -3.19 $ &  $1.66 $& $-1.53$\\
%\\
$J_\mu^2$ &$[dd] [sc] \bar c$& $ ~~3.19$ &  $ 1.66 $& ~~4.85\\
%\\
&$[ss] [uc] \bar c$& $ ~~0.00$&  $ 1.57$& ~~1.57\\
%\\
&$[ss] [dc] \bar c$& $~~3.16$ &  $ 1.64$& ~~4.80\\
%\\
	   \hline\hline
	   	  	  
\end{tabular}
\end{center}
%\end{ruledtabular}
\end{table}

\end{widetext}
 \begin{widetext}
 
  \begin{table}[htb!]
	\addtolength{\tabcolsep}{10pt}
	\caption{The contribution of light and heavy quarks to the magnetic octupole moment of the $P_{c(s)}$ states ($10^{-3}$\,fm$^3$).}
	\label{parameter4}
		\begin{center}
		%\begin{ruledtabular}
\begin{tabular}{l|ccccc}
	   \hline\hline
	   	   %%%%%%%%%%%%%%%%%%%%%%%%
	%   	  	   \\
   Current&Pentaquarks& $\mathcal O_{q}$& $\mathcal O_{c}$& $\mathcal O_{total}$\\
  % \\
     \hline\hline
     &$[uu] [dc] \bar c$& $ ~~0.24 $  &  $-0.80 $& $-0.56$\\
     %%%%
     &$[dd] [uc] \bar c$     &$~~0.00 $       &  $ -0.81$& $-0.81$\\
%\\
&$[uu] [sc] \bar c$& $ ~~0.26 $  &  $-0.84 $& $-0.58$\\
%\\
$J_\mu^1$ &$[dd] [sc] \bar c$     &$-0.25 $       &  $ -0.84$& $-1.09$\\
%\\
&$[ss] [uc] \bar c$& $~~0.00$    &  $-0.80$&$-0.80$\\
%\\
&$[ss] [dc] \bar c$& $-0.23$    &  $-0.81 $& $-1.04$\\
%\\
	   \hline\hline
	   %%%%%%%%%%%%%%%%%%%%%%%%%%%%%%%%%%%%%%%%%%%%%%%%%%%%%
	%   \\
	      Current&Pentaquarks& $\mathcal O_{q}$& $\mathcal O_{c}$& $\mathcal O_{total}$\\
  % \\
     \hline\hline
     &$[uu] [dc] \bar c$& $ ~~0.46 $   &  $-0.66 $& $-0.20$\\
&$[dd] [uc] \bar c$& $~~0.00$      &  $-0.62$&$-0.62$\\
%\\
&$[uu] [sc] \bar c$& $ ~~0.48 $   &  $-0.69 $& $-0.21$\\
%\\
$J_\mu^2$ &$[dd] [sc] \bar c$  & $ -0.43 $    &  $ -0.69$& $-1.12$\\
%\\
&$[ss] [uc] \bar c$& $~~0.00$      &  $-0.63$&$-0.63$\\
%\\
&$[ss] [dc] \bar c$& $-0.39$     &  $-0.67 $& $-1.06$\\
%\\
	   \hline\hline
\end{tabular}
\end{center}
%\end{ruledtabular}
\end{table}

\end{widetext}

%\section{Acknowledgments}
%The author acknowledges A. \"{O}zpineci for his valuable contributions to the comments, discussions, and suggestions.
%\vspace*{5 cm}
%\newpage

 %\vspace*{5 cm}
 % \newpage

 %\vspace*{5 cm}
  \newpage
  
  \begin{widetext}

  \appendix
  \section*{Appendix: The obtained sum rules for the electromagnetic multipole moments of $P_{c(s)}$ states for the ${J_\mu^1}$ current}\label{appenda}
 The obtained sum rules for the electromagnetic multipole moments of $P_{c(s)}$ statesfor the ${J_\mu^1}$ current are presented as follows,
 \begin{align}
  \label{edmn015}
&\mu^{J_\mu^1}_{ P_{c(s)}} \,\lambda^2_{ P_{c(s)}}=e^{\frac{m^2_{ P_{c(s)}}}{\rm{M^2}}}\, \rho_1 (\rm{M^2},\rm{s_0}),\\
%%%%%%%%%%%%%%%%%%%%%%
&\mathcal Q^{J_\mu^1}_{ P_{c(s)}} \,\lambda^2_{ P_{c(s)}}=e^{\frac{m^2_{ P_{c(s)}}}{\rm{M^2}}}\, \rho_2 (\rm{M^2},\rm{s_0}),\\
%%%%%%%%%%%%%%%%%%%%%%
&\mathcal O^{J_\mu^1}_{ P_{c(s)}} \,\lambda^2_{ P_{c(s)}}=e^{\frac{m^2_{ P_{c(s)}}}{\rm{M^2}}}\, \rho_3 (\rm{M^2},\rm{s_0}),
 \end{align}
with
\begin{align}
  \rho_1 (\mathrm{M^2},\mathrm{s_0}) &=  F_1(\mathrm{M^2},\mathrm{s_0}) -\frac{1}{m_{P_{c(s)}}} F_2(\mathrm{M^2},\mathrm{s_0}),\\
  %%%%%%%%%%%%%%%%%%%%%%%%%%%%%%%%%
  \rho_2 (\mathrm{M^2},\mathrm{s_0}) &=  F_1(\mathrm{M^2},\mathrm{s_0}) -m_{P_{c(s)}} F_3(\mathrm{M^2},\mathrm{s_0}),\\
  %%%%%%%%%%%%%%%%%%%%%%%%%%%%
  \rho_3 (\mathrm{M^2},\mathrm{s_0}) &=  F_1(\mathrm{M^2},\mathrm{s_0}) -\frac{1}{m_{P_{c(s)}}} F_2(\mathrm{M^2},\mathrm{s_0})-m_{P_{c(s)}} F_3(\mathrm{M^2},\mathrm{s_0})- 2 m^3_{P_{c(s)}} F_4(\mathrm{M^2},\mathrm{s_0}),
 \end{align}
where
\begin{align}
 F_1 (\mathrm{M^2},\mathrm{s_0})&= \frac{1}{2^{24}\times 3^3 \times 5^3 \times 7^2  \pi^7} \Bigg[6559  e_c m_c (2 m_{q_1} + m_{q_2})   I[0, 
   6] - 45 \Big(19 e_c + 26 (2 e_{q_1} + e_{q_2})\Big) I[0, 7]\Bigg]
    %%%%%%%%%%%%%%%%%%%%%%%%%%%%%
 \nonumber\\
 &+\frac{C_1 C_2 C_3}{2^{17}\times 3^6 \times 5 \times  \pi^3} \Bigg[ 15 \Big (3 m_c e_ {q_ 1} (m_ {q_ 1} + 4 m_ {q_ 2}) + 
    2 e_c m_c (7 m_ {q_ 1} + 4 m_ {q_ 2})\Big)  I[0, 1] - 
 22 (2 e_c + 3 e_ {q_ 1}) I[0, 2]\Bigg]
 %%%%%%%%%%%%%%%%%%%%%%%%%%%%%
 %%%%%%%%%%%%%%%%%%%%%%%%%%%%%
 \nonumber\\
 &+\frac{C_1 C_2^2}{2^{17}\times 3^6 \times 5   \pi^3} \Bigg[ 15 m_c\Big (20 e_c m_ {q_ 1} + 3 e_ {q_ 1} m_ {q_ 1} - 
    12 e_ {q_ 2} m_ {q_ 1} + 8 e_c m_ {q_ 2} + 
    12 e_ {q_ 1} m_ {q_ 2}\Big)  I[0, 1] - 
 22 (2 e_c  \nonumber\\
  &+ 3 e_ {q_ 1}) I[0, 2]\Bigg]
 %%%%%%%%%%%%%%%%%%%%%%%%%%%%%
 \nonumber\\
  &+\frac{C_1 C_2}{2^{23}\times 3^7 \times 5   \pi^5}
 \Bigg [-4 e_ {q_ 2} m_ 0^2 (1012 m_ {q_ 1} + 8553 m_c) I[0, 2] + 
 12 e_c \Big ( 
    m_ 0^2 ( 209 m_ {q_ 1} + 99 m_ {q_ 2} + 1326 m_c) I[0, 
          2])\Big) 
    \nonumber\\
  &  - 
 8 e_c (1109 m_ {q_ 1} + 877 m_ {q_ 2} + 1494 m_c) I[0, 3]+ 
 8 e_ {q_ 2} (2453 m_ {q_ 1}  + 3708 m_c) I[0, 3]
  + 
 e_ {q_ 1} \Big ( 
    m_ 0^2 ( (-5566 (m_ {q_ 1} + m_ {q_ 2})
 \nonumber\\
  & + 6591 m_c) I[0, 2]) 
         + 
    2 (6161 m_ {q_ 1} + 6296 m_ {q_ 2} - 456 m_c) I[0, 3]\Big)\Bigg]
      %%%%%%%%%%%%%%%%%%%%%%%%%%%%%
 \nonumber\\
  &+\frac{C_1 C_3}{2^{23}\times 3^7 \times 5   \pi^5}\Bigg [  
 12 e_c \Big ( 
    m_ 0^2 ( (-132 m_ {q_ 1} + 231 m_ {q_ 2} + 302 m_c)I[0, 
          2])\Big) - 
 8 ec \big (930 m_ {q_ 1} - 53 m_ {q_ 2}  + 484 m_c\big) I[0, 3]
          \nonumber\\
  &  + 
 e_ {q_ 1} \Big ( 
    m_ 0^2 (
     (-2376 m_ {q_ 1} - 3190 m_ {q_ 2} + 4263 m_c) I[0, 
         2]) + 4 (3436 m_ {q_ 1} - 288 m_ {q_ 2} - 105 m_c) I[0, 
       3]\Big)
 \Bigg]
      %%%%%%%%%%%%%%%%%%%%%%%%%%%%%
 \nonumber\\
 &
-\frac{ C_2^2\, e_c}{2^{14}\times 3^5 \times 5   \pi^3}
 \Bigg[ 145 m_ 0^2 m_ {q_ 1} m_c I[0, 2] + 
 12 (m_ {q_ 1} - 29 m_ {q_ 2}) m_c I[0, 3] + 9 I[0, 4]\Bigg] 
  %%%%%%%%%%%%%%%%%%%%%%%%%%%%
     \nonumber\\
      &+\frac{ C_1}{2^{28}\times 3^7 \times 5^2  \pi^7}
 \Bigg[ 105 \Big (  m_ {q_ 1} (-4 (482 e_ {q_ 1} + 
          5747 e_ {q_ 2}) - 2051 e_ {q_ 1}   m_ {q_ 2} + 
       8 e_c (365 m_ {q_ 1} + 346 m_ {q_ 2})) m_c \Big)  I[0, 4] \nonumber\\
 & + 
 4 \big (36876 e_c + 33455 e_ {q_ 1} - 42458 e_ {q_ 2}\big)  I[0, 5]\Bigg]
 %%%%%%%%%%%%%%%%%%%%%%%%%%%%
 \nonumber%\\
 \end{align}
\begin{align}
 &+\frac{1}{2^{21}\times 3^5 \times 5^2 \times 7 \pi^5}\Bigg[ - 
 15 \Big (459 m_ 0^2 ((e_ {q_ 1} + e_ {q_ 2}) m_ {q_ 1} C_ 2
 + 
       e_ {q_ 1} m_ {q_ 2} C_ 3)+ 
    7 e_c \big ( 
        m_ 0^2 (9 m_ {q_ 2} (16 C_ 2 - C_ 3) \nonumber\\
 &- 
            119 m_c (2 C_ 2 + C_ 3) + 
            18 m_ {q_ 1} (7 C_ 2 + 8 C_ 3))\big)\Big) I[0, 4] + 
 4 \Big (2 e_c (189 m_ {q_ 2} (7 C_ 2 - 3 C_ 3)- 
       664 m_c (2 C_ 2 + C_ 3) \nonumber\\
 &   + 189 m_ {q_ 1} (C_ 2 + 7 C_ 3))  + 
    81 (47 e_ {q_ 2} m_ {q_ 1} C_ 2  + 
        e_ {q_ 1} (-4 m_ {q_ 1} C_ 2+ 51 m_ {q_ 2} C_ 2 + 
           51 m_ {q_ 1} C_ 3 - 4 m_ {q_ 2} C_ 3))\Big) I[0, 5]\Bigg],
     %\nonumber
     \\
%\end{align} 
%
%\begin{align}
 F_2 (\mathrm{M^2},\mathrm{s_0})&= \frac{ m_c}{2^{24}\times 3^5 \times  5^3 \times 7^2   \pi^7} \Bigg[e_c \Big (12981 I[0, 7] + 
    280 \big ((- 
          148 (2 m_ {q_ 1} + m_ {q_ 2}) m_c) I[0, 6] \big) + 86226 I[1, 6]\Big)
        \nonumber\\
 &   -
 54 (2  e_ {q_ 1} + e_ {q_ 2})\Big (16  I[0, 7] \big) +
    427  I[1, 6]\Big)\Bigg]
 %%%%%%%%%%%%%%%%%%%%%%%%%%%%%%555
 \nonumber\\
 &+\frac{C_1 C_2 C_3\, m_c}{2^{17}\times 3^6  \pi^3} \Bigg[3 e_ {q_ 1} \Big (48 ( m_ {q_ 1} + m_ {q_ 2}) m_c I[0, 1] + 
    13 I[0, 2] - 62 I[1, 1]\Big) + 
 2 e_c \Big (2 (47  m_ {q_ 1} + 32 m_ {q_ 2}) m_c I[0, 1] \nonumber\\
 &- 
    17 I[0, 2] + 16 I[1, 1]\Big)
 \Bigg]
 %%%%%%%%%%%%%%%%%%%%%%%%%%%%%
 \nonumber\\
 &+\frac{C_1 C_2^2\, m_c}{2^{16}\times 3^6   \pi^3} \Bigg[  94 e_c m_ {q_ 1} m_c I[0, 1] - 576 e_ {q_ 2} m_ {q_ 1} m_c I[0, 1] + 
 11 e_c I[0, 2] + 102 e_ {q_ 2} I[0, 2] - 
 2 (5 e_c + 66 e_ {q_ 2}) I[1, 1]  \Bigg]
 %%%%%%%%%%%%%%%%%%%%%%%%%%%%%
 \nonumber\\
 &+\frac{C_1 C_2\, m_c}{2^{22}\times 3^7 \times 5   \pi^5}
 \Bigg [ 20 e_c \Big (-59 m_ {q_ 1} I[0, 3] - 
    2 (64 m_ {q_ 2} + 401 m_c) I[0, 3]  \nonumber\\
 &+ 
    3 m_ 0^2 (-3 (17 m_ {q_ 2} + 35 m_c) I[0, 2] + 
        4 m_ {q_ 1} (47 m_ {q_ 2} m_c I[0, 1] + 33 I[0, 2] ) + 24 (2 m_ {q_ 2} + 15 m_c) I[1, 1])\Big) \nonumber\\
 & + 
 6 e_c (801 m_ {q_ 1} - 40 (16 m_ {q_ 2} + 75 m_c)) I[1, 2] + 
 e_ {q_ 1}  \Big (40 (201 m_ {q_ 2} - 46 m_c) I[0, 3] + 
    45 m_ 0^2 (- 
       124 m_ {q_ 1} I[0, 2] \nonumber\\
 & + 78 m_ {q_ 2} I[0, 2] - 
       185 m_c I[0, 2] + 
       4 (31 m_ {q_ 1} - 93 m_ {q_ 2} + 89 m_c) I[1, 1]) + 
    2 m_ {q_ 1} (2816 I[0, 3] \nonumber\\
 & - 
       6480 m_ {q_ 2} m_c (I[0, 2] - 2 I[1, 1]) - 8409 I[1, 2]) + 
    7440 (3 m_ {q_ 2} - 2 m_c) I[1, 2]\Big) + 
 4 e_ {q_ 2} \Big (2 (371 m_ {q_ 1} + 1210 m_c) \nonumber\\
 & \times  I[0, 3]+ 
    45 m_ 0^2 ((116 m_ {q_ 1} + 325 m_c) I[0, 2] - 
       2 (88 m_ {q_ 1} + 305 m_c) I[1, 1]) + 
    3 (2809 m_ {q_ 1} + 7880 m_c) I[1, 2]\Big)  \Bigg]
      %%%%%%%%%%%%%%%%%%%%%%%%%%%%%
 \nonumber\\
   &+\frac{C_1 C_3\, m_c}{2^{22}\times 3^7 \times 5   \pi^5}\Bigg [ 2 e_ {q_ 1} \Big ((2327 m_ {q_ 2} + 370 m_c) I[0, 3] + 
    60 m_ {q_ 1} (67 I[0, 3] - 
       108 m_ {q_ 2} m_c (I[0, 2] - 2 I[1, 1])) 
       \nonumber\\
   &+ 
    45 m_ 0^2 ( 
        39 m_ {q_ 1} I[0, 2] - 62 m_ {q_ 2} I[0, 2] - 
        154 m_c I[0, 2] + 
        31 (-6 m_ {q_ 1} + 2 m_ {q_ 2} + 7 m_c) I[1, 1])\Big)\nonumber\\
   & + 
 3 e_ {q_ 1} (7440 m_ {q_ 1} - 5117 m_ {q_ 2} - 6520 m_c) I[1, 2] + 
 4 e_c \Big (-2 (18 m_ {q_ 2} + 665 m_c) I[0, 3] + 
    15 m_ 0^2 ( - 
       51 m_ {q_ 1} I[0, 2]
    \nonumber\\
   & + 30 m_ {q_ 2} I[0, 2] - 
       102 m_c I[0, 2] + 
       4 (12 m_ {q_ 1} - 5 m_ {q_ 2} + 36 m_c) I[1, 1]) + 
    6 (53 m_ {q_ 2} - 300 m_c) I[1, 2] \nonumber\\
   & - 
    20 m_ {q_ 1} (32 I[0, 3] + 
        27 m_ {q_ 2} m_c (3 I[0, 2] - 8 I[1, 1]) + 48 I[1, 2])\Big)\Bigg]
      %%%%%%%%%%%%%%%%%%%%%%%%%%%%%
 \nonumber  \\
 &-\frac{ C_2^2\,m_c}{2^{14}\times 3^5 \times 5  \pi^3}
 \Bigg[  9 e_ {q_ 2} (I[0, 4] - 4 I[1, 3]) + 
 e_c \Big (4 (
       10 (m_ {q_ 1} - 9 m_ {q_ 2}) m_c) I[0, 3] - 15 I[0, 4] + 
    2 m_ 0^2 (75 m_ {q_ 1} m_c 
    \nonumber\\
 & \times I[0, 2] + 47 (I[0, 3] - 3 I[1, 2]))  + 176 I[1, 3]\Big)\Bigg] 
  %%%%%%%%%%%%%%%%%%%%%%%%%%%%
     \nonumber\\
          &+\frac{ C_1\, m_c}{2^{26}\times 3^6 \times 5^2    \pi^7}
 \Bigg[ 891 e_c I[0, 5] + 
 160 e_c \Big (( 295 m_ {q_ 1} m_ {c} + 
       113 m_ {q_ 2} m_ {c}) I[0, 4] + 
    2 ( + 105 m_ {q_ 1} m_ {c}   + 
       42 m_ {q_ 2} m_ {c}) \nonumber\\
     & \times I[1, 3]\Big) - 6630 e_c I[1, 4] - 
 10 e_ {q_ 2} \Big (132 I[0, 5] + 
    176 m_ {q_ 1} m_ {c} (37 I[0, 4] + 104 I[1, 3]) + 
    355 I[1, 4]\Big) \nonumber\\
     &- 
 4 e_ {q_ 1} \Big ( - 
       286 m_ {q_ 1} m_ {c} - 235 m_ {q_ 2} m_ {c}) I[0, 4] + 
    453 I[0, 5] + 
    80 (- 97 m_ {q_ 1} m_ {c} 
     - 
       136 m_ {q_ 2} m_ {c}) I[1, 3] + 4295 I[1, 4]\Big) \Bigg]
 %%%%%%%%%%%%%%%%%%%%%%%%%%%%
 \nonumber%\\
       \end{align}
\begin{align}
 &+\frac{m_c}{2^{20}\times 3^4 \times 5^2   \pi^5}\Bigg[12 \bigg (2 e_ {q_ 2} m_ {q_ 1} C_ 2 (6 I[0, 5] + 
       5 m_ 0^2 (I[0, 4] - 8 I[1, 3])) + 
    e_ {q_ 1} \Big (3 (3 m_ {q_ 1} C_ 2 + m_ {q_ 2} C_ 2 + 
           m_ {q_ 1} C_ 3  \nonumber\\
 & + 3 m_ {q_ 2} C_ 3) I[0, 5] + 
        10 m_ 0^2 \big ((-2 m_ {q_ 1} C_ 2 + 3 m_ {q_ 2} C_ 2 + 
               3 m_ {q_ 1} C_ 3 - 2 m_ {q_ 2} C_ 3) I[0, 4] + 
            4 (m_ {q_ 1} C_ 2 - 3 m_ {q_ 2} C_ 2  \nonumber\\
 &- 3 m_ {q_ 1} C_ 3 + 
               m_ {q_ 2} C_ 3)  I[1, 3]\big)\Big)\bigg) + 
 630 \Big (e_ {q_ 2} m_ {q_ 1} C_ 2 - 
    e_ {q_ 1} (m_ {q_ 1} (C_ 2 - 2 C_ 3) + 
        m_ {q_ 2} (-2 C_ 2 + C_ 3))\Big) I[1, 4] 
 \nonumber\\
 &+ 
 e_c \bigg (-(-92 m_c (2 C_ 2 + C_ 3)+ 
        m_ {q_ 2} (224 C_ 2 + 99 C_ 3) + 
        m_ {q_ 1} (422 C_ 2 + 224 C_ 3)) I[0, 5] + 
    20 m_ 0^2 \Big ((m_ {q_ 1} (52 C_ 2  \nonumber\\
 &- 30 C_ 3) - 
          12 m_c (2 C_ 2 + C_ 3) + m_ {q_ 2} (-30 C_ 2+ 41 C_ 3)) I[
         0, 4] + 176 (2 m_ {q_ 2} C_ 2 + 2 m_ {q_ 1} C_ 3 - 
          m_ {q_ 2} C_ 3) I[1, 3]\Big) \bigg)\Bigg],\\
       %%%%%%%%%%%%%%%%%%%%%%%%%%%%%%
%\end{align}
%
%\begin{align}
 F_3 (\mathrm{M^2},\mathrm{s_0})&= \frac{ m_c}{2^{22}\times 3^4 \times  5^3 \times 7   \pi^7}
 \Bigg[ e_c \Big (3280 m_c (2 m_ {q_ 1} + m_ {q_ 2})  I[0, 5] - 
    2387 I[0, 6] - 2208 I[1, 5]\Big) + 
 9 (2  e_ {q_ 1} + e_ {q_ 2}) \nonumber\\
 & \times (122 I[0, 6] + 81 I[1, 5])\Bigg]
 %%%%%%%%%%%%%%%%%%%%%%%%%%%%%%%%%%%%
 \nonumber\\
 &+\frac{C_1 C_2\, m_c}{2^{20}\times 3^6 \times 5   \pi^5}
 \Bigg [ 3 m_ {q_ 1} \Big (-490 e_ {q_ 1} m_ 0^2 I[0, 1] - 
    400 e_ {q_ 2} m_ 0^2 I[0, 1] + 953 e_ {q_ 1} I[0, 2] + 
    706 e_ {q_ 2} I[0, 
      2] + (365 e_ {q_ 1} \nonumber\\
 &+ 232 e_ {q_ 2}) I[1, 1]\Big) + 
 2 e_c \Big (20 m_ 0^2 (7 m_ {q_ 1} - 30 m_c) I[0, 1] - 
    309 m_ {q_ 1} I[0, 2] + 440 m_c I[0, 2] + 
    20 m_ {q_ 1} I[1, 1]\Big)\Bigg]
    %%%%%%%%%%%%%%%%%%%%%%%%%%%%%%%%%%%%
 \nonumber\\
 &+\frac{C_1 C_3\, m_c}{2^{21}\times 3^6 \times 5   \pi^5}
 \Bigg [ 3 e_ {q_ 1} m_ {q_ 2} (-980 m_ 0^2 I[0, 1] + 1873 I[0, 2] + 
    706 I[1, 1]) + 
 8 e_c \Big (10 m_ 0^2 (5 m_ {q_ 2} - 33 m_c) I[0, 1] \nonumber\\
 &- 
    129 m_ {q_ 2} I[0, 2] + 260 m_c I[0, 2] - 
    28 m_ {q_ 2} I[1, 1]\Big)\Bigg]\nonumber\\
    %%%%%%%%%%%%%%%%%%%%%%%%%%%%%%%%%%%%%%%%%%%%%%%%%%%%%%%%%%%%%%%%%%%%%%%%%%%%%%%%%%%%%%%%%%%%%%%%%%%%%%%%%%%
    &-\frac{ C_2^2}{2^{11}\times 3^4 \pi^3}
 \Bigg[ e_c m_{q_1} m_c^2 (5 m_0^2 I[0, 1] - 4 I[0, 2])  \Bigg] 
  %%%%%%%%%%%%%%%%%%%%%%%%%%%%
     \nonumber\\
     &+\frac{ C_1\, m_c}{2^{25}\times 3^6 \times 5 \times 7    \pi^7}
 \Bigg[ -896 e_c m_c (13 m_ {q_ 1} + 17 m_ {q_ 2})  I[0, 3] + 
 21 (-161 e_c + 411 e_ {q_ 1} + 70 e_ {q_ 2}) I[0, 4]  \nonumber\\
     &+ 
 2 (-764 e_c+ 2069 e_ {q_ 1} + 270 e_ {q_ 2}) I[1, 3] \Bigg]
 %%%%%%%%%%%%%%%%%%%%%%%%%%%%
 \nonumber\\
  &+\frac{m_c}{2^{19}\times 3^4 \times 5 \times 7   \pi^5}\Bigg[  -42 ((e_ {q_ 1} + e_ {q_ 2}) m_ {q_ 1} C_ 2 + 
    e_{q_1} m_ {q_ 2} C_ 3) (-21 I[0, 4] + 
    m_ 0^2 (16 I[0, 3] + 21 I[1, 2])) \nonumber\\
    %%%%%%%%%%%%%%%%%%%%%%%%%%%%
     &+ 
 7 e_c \Big (-3 (178 m_ {q_ 1} C_ 2 + 89 m_ {q_ 2} C_ 3 + 
       12 m_c (2 C_ 2 + C_ 3)) I[0, 4] + 
    8 m_ 0^2 \big (4 (26 m_ {q_ 1} C_ 2 + 8 m_c C_ 2 + 
           13 m_ {q_ 2} C_ 3 \nonumber\\
 &+ 4 m_c C_ 3) I[0, 3] + 
        57 (2 m_ {q_ 1} C_ 2 + m_ {q_ 2} C_ 3) I[1, 2]\big)\Big) + 
 756 (e_ {q_ 1} + e_ {q_ 2}) m_ {q_ 1} C_ 2 I[1, 3] - 
 1856 e_c (2 m_ {q_ 1} C_ 2 \nonumber\\
 &+ m_ {q_ 2} C_ 3) I[1, 3]\Bigg],\\
 %%%%%%%%%%%%%%%%%%%%%%%%%%%%%%%%%%%%%%%%
  F_4 (\mathrm{M^2},\mathrm{s_0})&= \frac{ m_c}{2^{22}\times 3^5 \times  5^3 \times 7   \pi^7}
 \Bigg[  \big(736 e_c - 243 (2 e_{q_1} + e_{q_2}) \big)I[0, 5]\Bigg]
 %%%%%%%%%%%%%%%%%%%%%%%%%%%%%%%%%%%%
 \nonumber\\
 &-\frac{C_1 C_2\, m_c\,m_{q_1}}{2^{19}\times 3^6 \times 5   \pi^5}
 \Bigg [ (40 e_c + 1095 e_{q_1} + 696 e_{q_2}) I[0, 1] \Bigg]
    %%%%%%%%%%%%%%%%%%%%%%%%%%%%%%%%%%%%
 \nonumber\\
 &+\frac{C_1 C_3\, m_c\,m_{q_2}}{2^{19}\times 3^6 \times 5   \pi^5}
 \Bigg [ (112 e_c - 1059 e_{q_1})  I[0, 1] \Bigg] 
  %%%%%%%%%%%%%%%%%%%%%%%%%%%%
     \nonumber%\\
\end{align}

\begin{align}
%\end{align}
%
%\begin{align}
     &+\frac{ C_1\, m_c}{2^{23}\times 3^6 \times 5 \times 7    \pi^7}
 \Bigg[ (764 e_c - 2069 e_{q_1} - 270 e_{q_2}) I[0, 3]\Bigg]
 %%%%%%%%%%%%%%%%%%%%%%%%%%%%
 \nonumber\\
  &+\frac{m_c}{2^{17}\times 3^4 \times 5 \times 7   \pi^5}\Bigg[ -4 e_c m_c (2 m_ {q_ 1} C_ 2 + m_ {q_ 2} C_ 3) (399 m_ 0^2 I[0, 2] - 
    232 I[0, 3]) + 
 63 m_c ((e_ {q_ 1} + e_ {q_ 2}) m_ {q_ 1} C_ 2\nonumber\\
  & + 
    e_ {q_ 1} m_ {q_ 2} C_ 3) (7 m_ 0^2 I[0, 2] - 6 I[0, 3])  \Bigg],
\end{align}

 \noindent  where $C_1 =\langle g_s^2 G^2\rangle$ is gluon condensate;  $C_2 =\langle \bar q_1 q_1 \rangle$ and $C_3 =\langle \bar q_2 q_2 \rangle$ are corresponding light-quark condensates.  For completeness, the values $e_{q_1}$, $e_{q_2}$,  $m_{q_1}$, $m_{q_2}$, $\langle \bar q_1 q_1 \rangle$, and $\langle \bar q_2 q_2 \rangle$  related to the expressions of the sum rules in aforementioned equations are given in Table \ref{eqimqi}. The function $\mathrm{I}[n,m]$ is given as 
\begin{align}
 \mathrm{I}[n,m]&= \int_{\widetilde m}^{\mathrm{s_0}} ds~ e^{-s/\mathrm{M^2}}~
 s^n\,(s-\widetilde m)^m,
 %I_1[\mathcal{F}]&=\int D_{\alpha_i} \int_0^1 dv~ \mathcal{F}(\alpha_{\bar q},\alpha_q,\alpha_g) \delta'(\alpha_ q +\bar v \alpha_g-u_0),\nonumber\\
 % I_2[\mathcal{F}]&=\int D_{\alpha_i} \int_0^1 dv~ \mathcal{F}(\alpha_{\bar q},\alpha_q,\alpha_g) \delta'(\alpha_{\bar q}+ v \alpha_g-u_0), \nonumber\\
 %\end{align}
 %\begin{align}
 %   I_3[\mathcal{F}]&=\int D_{\alpha_i} \int_0^1 dv~ \mathcal{F}(\alpha_{\bar q},\alpha_q,\alpha_g) \delta(\alpha_ q +\bar v \alpha_g-u_0),\nonumber\\
%   I_4[\mathcal{F}]&=\int D_{\alpha_i} \int_0^1 dv~ \mathcal{F}(\alpha_{\bar q},\alpha_q,\alpha_g)\delta(\alpha_{\bar q}+ v \alpha_g-u_0),\nonumber\\
%    I_5[\mathcal{F}]&=\int_0^1 du~ \mathcal{F}(u)\delta'(u-u_0),\nonumber\\
% I_6[\mathcal{F}]&=\int_0^1 du~ \mathcal{F}(u),
 \end{align}
 
 \end{widetext}
 
   \begin{widetext}

 \noindent where  
  $\widetilde m= (2m_c)^2$ for the $[u u][d c] \bar c$ and $[dd ][u c] \bar c$ states; 
 $\widetilde m= (m_s+2m_c)^2$ for the $[u u][s c] \bar c$ and $[dd ][s c] \bar c$ states; and, 
 $\widetilde m= (2m_s+2m_c)^2$ for the $[s s][u c] \bar c$ and $[s s][d c] \bar c$ states.
 \end{widetext}

   \begin{widetext}
   
    \begin{table}[htp]
	\addtolength{\tabcolsep}{10pt}
	\caption{ The values $e_{q_1}$, $e_{q_2}$,  $m_{q_1}$, $m_{q_2}$, $\langle \bar q_1 q_1 \rangle$, and $\langle \bar q_2 q_2 \rangle$ related to the expressions of the sum rules in aforementioned equations.}
	\label{eqimqi}
		\begin{center}
		\scalebox{0.85}{
\begin{tabular}{l|ccccccc}
                \hline\hline
              \\
 Parameters  & $[u u][d c] \bar c$  &  $[dd ][u c] \bar c$ & $[u u][s c] \bar c$ & $[dd ][s c] \bar c$ &  $[s s][u c] \bar c$ & $[s s][d c] \bar c$\\
     \\
                \hline\hline
   %                                   \\
 $e_{q_{1}}$& $e_u$ & $e_d$ & $ e_u $ &$ e_d $ & $ e_s $ & $ e_s $\\
  $e_{q_{2}}$&$e_d$ & $e_u$ & $ e_s $ & $ e_s $ & $ e_u $ & $ e_d $\\
   $m_{q_{1}}$&$m_u$ & $m_d$ & $m_u$  & $m_d$  & $m_s$ & $m_s$ \\
    $m_{q_{2}}$& $m_d$ & $m_u$ & $m_s$ & $m_s$ & $m_u$ & $m_d$\\
$\langle \bar q_1 q_1 \rangle$ &$\langle \bar uu \rangle$ & $ \langle \bar d d \rangle  $ & $\langle \bar uu \rangle$ & $\langle \bar dd \rangle$ & $\langle \bar ss \rangle$ & $\langle \bar ss \rangle$\\
$\langle \bar q_2 q_2 \rangle$ &$\langle \bar d d \rangle $ & $\langle \bar uu \rangle $ &$\langle \bar ss\rangle$& $\langle \bar ss \rangle$ & $\langle \bar uu \rangle$ & $\langle \bar dd \rangle$\\
    % \\
% \\
     \hline\hline
 \end{tabular}
}
\end{center}
\end{table}

 \end{widetext}
  
\bibliographystyle{elsarticle-num}
\bibliography{PccbarstatesMMrevised.bib}
\end{document}